\documentclass[sigconf]{acmart}

\AtBeginDocument{%
  }

\usepackage{bbding,pifont}
\usepackage{enumitem}
\usepackage{multirow}

\newcommand{\redHigh}{\textcolor{red}}
\newcommand{\blueHigh}{\textcolor{blue}}
\newcommand{\greenHigh}{\textcolor[HTML]{5EC265}}
\newcommand{\purpleHigh}{\textcolor[HTML]{9538B6}}

\newcommand{\prefix}{\textcolor[RGB]{0, 102, 204}}

\copyrightyear{2025}
\acmYear{2025}
\acmConference[IUI '25]{30th International Conference on Intelligent User Interfaces}{March 24--27, 2025}{Cagliari, Italy}
\acmBooktitle{30th International Conference on Intelligent User Interfaces (IUI '25), March 24--27, 2025, Cagliari, Italy}\acmDOI{10.1145/3708359.3712088}
\acmISBN{979-8-4007-1306-4/25/03}

\begin{document}


\title[StratIncon Detector]{StratIncon Detector: Analyzing Strategy Inconsistencies Between Real-Time Strategy and Preferred Professional Strategy in MOBA Esports}


\author{Ruofei Ma}
\email{marf2023@shanghaitech.edu.cn}
\orcid{0009-0005-0841-7251}
\affiliation{%
  \institution{ShanghaiTech University}
  \city{Shanghai}
  \country{China}
}

\author{Yu Zhao}
\email{zhaoyu2023@shanghaitech.edu.cn}
\orcid{0009-0001-9891-8893}
\affiliation{%
  \institution{ShanghaiTech University}
  \city{Shanghai}
  \country{China}
}

\author{Yuheng Shao}
\email{shaoyh2024@shanghaitech.edu.cn}
\orcid{0009-0008-6991-6427}
\affiliation{%
  \institution{ShanghaiTech University}
  \city{Shanghai}
  \country{China}
}

\author{Yunjie Yao}
\email{yaoyj2024@shanghaitech.edu.cn}
\orcid{0009-0000-8371-6984}
\affiliation{%
  \institution{ShanghaiTech University}
  \city{Shanghai}
  \country{China}
}

\author{Quan Li}
\authornote{Corresponding Author.}
\email{liquan@shanghaitech.edu.cn}
\orcid{0000-0003-2249-0728}
\affiliation{%
  \institution{ShanghaiTech University}
  \city{Shanghai}
  \country{China}
}








\renewcommand{\shortauthors}{Ma et al.}



\begin{abstract}
MOBA (Multiplayer Online Battle Arena) games require a delicate interplay of strategic planning and real-time decision-making, particularly in professional esports, where players exhibit varying levels of skill and strategic insight. While team strategies have been widely studied, analyzing inconsistencies in professional matches remains a significant challenge. The complexity lies in defining and quantifying the difference between real-time and preferred professional strategies, as well as understanding the disparities between them. Establishing direct causal links between specific strategic decisions and game outcomes also demands a comprehensive analysis of the entire match progression. To tackle these challenges, we present the \textit{StratIncon Detector}, a visual analytics system designed to assist professional players and coaches in efficiently identifying strategic inconsistencies. The system detects real-time strategies, predicts preferred professional strategies, extracts relevant human factors, and uncovers their impact on subsequent game phases. Findings from a case study, a user study with 24 participants, and expert interviews suggest that, compared to traditional methods, the \textit{StratIncon Detector} enables users to more comprehensively and efficiently identify inconsistencies, infer their causes, evaluate their effects on subsequent game outcomes, and gain deeper insights into team collaboration—ultimately enhancing future teamwork.
\end{abstract}






\maketitle

\section{Introduction}

\par MOBA (Multiplayer Online Battle Arena) games are characterized by team-based strategic play, requiring both complex planning and real-time decision-making~\cite{zhang2019hierarchical,ye2020supervised}. Their appeal lies in intricate mechanics and the depth of strategic thought involved, with effective strategies often determining match outcomes~\cite{ye2020supervised}. 

\par Players are typically classified as \textit{amateurs} or \textit{professionals}. Professionals engage in extensive strategic analysis, developing new team compositions and tactics, while amateurs play primarily for enjoyment~\cite{hodge2019win}. Professionals exhibit advanced skills through specialized training in teamwork and quick decision-making~\cite{hodge2019win}. They also maintain composure under pressure, resulting in decisions less swayed by emotions~\cite{trotter2021social}. Furthermore, professional matches tend to feature fewer errors, as players adopt more cautious strategies~\cite{liu2023bpcoach}. Consequently, our analysis prioritizes professional competitions to provide a comprehensive understanding of strategy deployment in high-stakes environments, minimizing biases from human factors.

\par In professional esports, match preparation and execution involve a complex and disciplined process. Players undergo rigorous training, rehearsing various scenarios and developing specific strategies for optimal outcomes~\cite{kari2016athletes}. Communication with coaches indicates that teams analyze both their own scrimmages and top-tier opponents' replays to adopt effective tactics, culminating in what we term the \underline{\textit{preferred professional strategy}}. This strategy encompasses rehearsed tactics and counter-strategies rather than a purely theoretical ideal~\cite{kari2016athletes}. Despite thorough preparations, the dynamic nature of competition introduces unforeseeable variables, requiring players to rely on their instantaneous judgment for real-time decision-making, referred to as \underline{\textit{real-time strategy}}. This approach emphasizes adaptability over rigid adherence to pre-established plans. However, discrepancies often arise between the real-time strategy employed and the preferred professional strategy~\cite{montag2022investigating}. These inconsistencies can stem from several factors: First, \textit{uncertainty about the opponent's strategy} may force players to deviate from the optimal plan~\cite{montag2022investigating}. Second, \textit{divergent interpretations of lineup and tactics} across teams result in varied strategic approaches~\cite{chen2021heroes,liu2023bpcoach}. Third, \textit{team communication and coordination} directly impact the execution of strategy, where poor communication can lead to deviations from the intended strategy~\cite{montag2022investigating}. Lastly, \textit{subjective factors}, such as psychological pressure, attention focus, experience, and intuition, play a significant role in decision-making under the high-pressure conditions of competitive play, potentially causing misalignment with the preferred professional strategy~\cite{trotter2021social}.

\par Examining the consistency between real-time strategy and preferred professional strategy is essential in professional esports~\cite{gisbert2024key,smith2021identifying,freeman2019understanding}. This analysis allows for a comprehensive review of the performance of the matches, allowing teams to refine strategies and identify key timing opportunities that control the tempo of the game. Addressing strategic inconsistencies improves the immediate outcomes of matches and contributes to long-term player development, as consistency and adaptability significantly impact market value and transfer prospects~\cite{zhong2022impact,nagorsky2020structure}. Experts indicate that clubs often prioritize players who demonstrate strategic consistency, teamwork, and effective execution qualities vital in an environment where skill parity is common. Moreover, analyzing strategic consistency is crucial for optimizing team collaboration and resource allocation. By understanding strategy alignment, teams can enhance synergy and efficiency during gameplay. Additionally, inconsistencies can reveal players' psychological states; recognizing discrepancies between real-time and preferred professional strategies may lead to frustration or anxiety, affecting future performance. Thus, exploring strategy inconsistencies is vital for improving strategy analysis, fostering team collaboration, and understanding player psychology, ultimately influencing the development of both players and teams.
\par Previous research on team strategies in esports has often combined qualitative and quantitative methods to provide a holistic analysis~\cite{chen2021heroes}. This work typically focuses on pre-match lineup decisions~\cite{chen2021heroes} and broader strategic planning~\cite{young2012evolutionary,gao2021learning}. Some studies delve into in-game strategies~\cite{cavadenti2016did,drachen2014skill,rioult2014mining,yang2014identifying,ahmad2019modeling}, using spatial-temporal data modeling to simulate real-time strategies among average players~\cite{ahmad2019modeling}, and employing visualization tools to enhance model interpretability~\cite{vsufliarsky2023through}. Despite these efforts, significant challenges remain in analyzing strategy inconsistencies in professional matches, particularly in three key areas: \textit{First, defining and quantifying real-time versus preferred professional strategies is complicated by the dynamic nature of MOBA games~\cite{jin2023actorlens}}. Most studies evaluate strategy effectiveness through post-match metrics—like outcomes and performance statistics—yet these often overlook the nuances of strategy definition~\cite{jin2023actorlens,yang2022predicting,aung2018predicting}. The lack of comprehensive log data further complicates this analysis, as does the challenge of handling large, multidimensional datasets that reflect variables such as player actions and resource management. 
\textit{Second, understanding why players deviate from preferred professional strategies requires analyzing both the strategies themselves and the contextual factors influencing these decisions.} In-game conditions and individual player styles play crucial roles, and personal skills, psychological states, and team dynamics contribute to these deviations. The preferred professional strategy may be theoretically ideal but often does not align with individual player preferences or interpretations of similar in-game situations.
\textit{Third, measuring the impact of strategic inconsistencies on match outcomes is complex, as factors like player skill, teamwork, and opponent strategies interact simultaneously.} This makes isolating the effects of specific inconsistencies challenging~\cite{jones2016cumulative}. For example, if the Top Laner and Mid Laner of a team exhibit strategic inconsistencies leading to the loss of a turret, the outcome could result from multiple influences. Therefore, evaluating long-term effects necessitates a comprehensive analysis of the entire match progression.


\par To tackle these intricate challenges, we explored five research questions (\textbf{RQ1-RQ5}). First, we examined factors influencing professional players' decision-making and their challenges in identifying strategy inconsistencies, leading to \textbf{RQ1: What are the key considerations for decision-making?} and \textbf{RQ2: What methods are acceptable for inconsistencies analysis?} To answer RQ1, we analyzed a professional team's post-game data using contextual inquiry, identifying five findings on traditional analysis methods and conducting Phase I interviews with four experts to pinpoint three key decision-making considerations. For RQ2, we proceeded with Phase II of interviews, deriving four design requirements, guiding the iterative development of a visualization system. Building on this prototype, we explored \textbf{RQ3: How do we evaluate usability and effectiveness of our methods?} \textbf{RQ4: What characterizes user interaction with the system?} and \textbf{RQ5: How do players trust and collaborate with the system?} To address these questions, we conducted a case study on key in-game events, complemented by expert interviews and a user study. The case study illustrated how experts engaged with the system to refine their decisions iteratively. Expert interviews yielded insights into \textit{system performance}, \textit{visual design}, \textit{interaction}, \textit{generalizability}, and \textit{scalability}. Additionally, a between-subjects user study with $24$ participants—comprising players, coaches, and analysts from university esports teams—highlighted the system's strengths in insightfulness, system performance, user experience, and team collaboration. The key contributions of this study are summarized as follows:

\begin{itemize}
\item We gained valuable insights into strategy analysis and decision-making by conducting a Contextual Inquiry during a professional esports team's post-game analysis and interviewing various players.
\item We quantified real-time and preferred professional strategies using dynamic sliding window prediction on in-game data, blending historical performance and preferences. Additionally, we developed the visual analytics system \textit{StratIncon Detector} for in-depth analysis of strategic discrepancies.
\item We validated the backend model's effectiveness through benchmark experiments and assessed the system's efficacy via user studies, case studies, and expert feedback.
\end{itemize}

\section{Background and Related Work}
\subsection{Game Mechanics}
\par MOBA (Multiplayer Online Battle Arena) is a genre of games that focuses on team cooperation and strategy. Players control unique characters with distinct abilities, and the primary objective is to destroy the enemy's main base. By killing enemies and controlling resources, players enhance their characters' abilities to ultimately secure victory. One of the most popular examples of a MOBA game is League of Legends (LOL). In the game, players are divided into two teams of five (The Top Laner, Jungler, Mid Laner, Bot Laner, Support), with each player controlling a ``champion'' that has unique skills and attributes. The map is divided into three main lanes (Top, Mid, and Bot) and a jungle area. Players accumulate gold and experience by killing minions, enemy champions, and jungle monsters, gradually strengthening their champions. The ultimate goal is to destroy the enemy's nexus. For more details on the basic terminology and metrics of MOBA games, please refer to \autoref{appendix:moba_fundamental}.
\subsection{Team Strategies Analysis}
\par Data analytics and AI have significantly advanced team strategy analysis in competitive sports, particularly in esports~\cite{tian2019use,wu2018forvizor,yang2022predicting,chen2021heroes, liu2023bpcoach,robertson2021wait,penney2021shoutcasters,cheung2011starcraft}. In \textbf{holistic strategy analysis}, substantial work has focused on understanding how team composition affects game outcomes~\cite{liu2023bpcoach,chen2021heroes}. For example, Chen et al.~\cite{chen2021heroes} used statistical models and machine learning to identify patterns in lineup selection, examining how different character combinations influence success. Liu et al.~\cite{liu2023bpcoach} introduced a lineup selection visual analytics system based on Monte Carlo tree search to assist professional teams in making strategic lineup choices. 
Other studies have concentrated on \textbf{in-game strategies}~\cite{cavadenti2016did,drachen2014skill,rioult2014mining}, such as Rioult et al.~\cite{rioult2014mining}, who analyzed player and team tactics through topological measures to predict match outcomes.
Previous research has also focused on designing more sophisticated AI agents~\cite{ahmad2019modeling,zhang2019hierarchical,ye2020supervised}. Ahmad et al.~\cite{ahmad2019modeling} differentiated between macro-strategy and micro-manipulation, using different learning methods to improve agent behavior. Several studies~\cite{ye2020mastering,ye2020supervised,zhang2019hierarchical} demonstrated that top-tier AI agents could match high-ranking human players, benefiting professional training. However, these advanced strategies often prove difficult for human players to understand and learn from. Additionally, many studies require pre-identified labels for decision-making and tactical behaviors~\cite{zhang2019hierarchical,ye2020supervised}, typically based on traditional events like first blood or first turret, limiting strategic analysis depth. Our work addresses these challenges by predicting strategy triads—(who, where, what)—to uncover strategic inconsistencies at a more granular level. Unlike studies focused on average players, we examine professional competitions, taking into account the unique differences among elite players. Beyond identification, we emphasize the interpretability of strategic inconsistencies, aiming to provide professional players with insights to enhance their understanding and facilitate continuous improvement in their gameplay strategies.

\par As team strategy analysis advances, team collaboration and conflict management have become critical areas of focus. Several studies~\cite{wieland2013turning,shaikh2024rehearsal,whiting2019did,aritzeta2005team,weingart2023manage} have explored how to effectively resolve internal conflicts and improve collaboration, thereby enhancing team performance. For example, Shaikh et al.~\cite{shaikh2024rehearsal} developed an interactive system, which uses simulated dialogues based on the Interest-Rights-Power (IRP) theory to help users practice conflict resolution strategies in a virtual environment. The study showed that using this system significantly reduced the frequency of confrontational strategies in actual conflicts while increasing the use of collaborative strategies. Additionally, Whiting et al.~\cite{whiting2019did} found that, during the early stages of team collaboration, designing friendly interactions or providing a positive starting environment could help reduce the risk of team fractures. Building on these findings, our proposed inconsistency analysis system presents all potential conflicts through an interactive interface for team discussion. This approach helps identify strategic issues, fosters collaboration, and reduces the risk of internal conflicts.

\subsection{Time Series Data Analysis and Forecasting}
\par MOBA games, with their dynamic, moment-to-moment nature, naturally generate time series data that offer valuable insights into gameplay and player behavior. Time series analysis and forecasting, particularly through sliding window techniques~\cite{hota2017time}, have been widely applied across various domains, including financial trading~\cite{conlon2009cross,sezer2019financial}, energy management~\cite{vafaeipour2014application,chou2016time}, and e-commerce~\cite{bandara2019sales,shao2019synchronizing}. This method, which segments data into overlapping or non-overlapping windows, allows for detailed temporal pattern analysis. In the automotive industry, Mozaffari et al.~\cite{mozaffari2015vehicle} utilized sliding window analysis to predict vehicle speeds, improving fuel efficiency and emission reduction. Similarly, Chou et al.~\cite{chou2016time} combined SARIMA with a MetaFA-LSVR model, employing a sliding window approach to forecast daily energy consumption. Doucoure et al.~\cite{doucoure2016time} leveraged neural networks to predict wind speed, enhancing wind power reliability. Building on these methods, our work applies sliding window-based time series forecasting to MOBA games, specifically to detect strategic inconsistencies. This novel approach brings predictive analytics to esports, creating new opportunities for performance insights.

\subsection{Visualization of Game Events Sequence}
\par Player behavior visualization, especially within game event sequences, has become a growing focus in gaming research~\cite{li2019visualizing,agarwal2020bombalytics,chen2017gamelifevis,afonso2019comparison,gonccalves2018analysing}. Early studies relied on timeline-based visualizations~\cite{wongsuphasawat2011lifeflow,li2019visualizing}, which were intuitive and aligned with user habits. For instance, Li et al.~\cite{li2019visualizing} developed a system to analyze player behaviors in puzzle games by visualizing event sequences. However, this approach assumes uniformity in scenarios, limiting its ability to capture complex behaviors. Agarwal et al.~\cite{agarwal2020bombalytics} introduced \textit{Bombalytics} to study competitive and cooperative strategies in team games, but it simplifies event sequences, making it less suitable for complex games with numerous neutral events, like MOBA games. Research has also expanded to spatiotemporal data and map-based visualizations~\cite{afonso2019comparison,afonso2021visualeague, gonccalves2018analysing,vsufliarsky2023through}. For example, the \textit{VisuaLeague} series~\cite{afonso2021visualeague} used dynamic maps and replay systems for visualizing matches but focused primarily on post-match summaries and lacked depth in event exploration and real-time analytics. Adam et al.~\cite{vsufliarsky2023through} addressed these gaps with the \textit{Spatio-Temporal Cube (STC)}, integrating player movements and events into a cube-based system to streamline visualization. Building on these approaches, our study combines inconsistency visualization with timelines to thoroughly explore strategic inconsistencies and evaluate their impact on in-game events.

\section{Observational Study}
\label{sec:observational study}

\subsection{Participants}
\par To understand the methodologies and challenges professional players and coaches face in managing strategy inconsistencies, we collaborated with a successful MOBA esports team that won a provincial league and ranked in the top four nationally. Our primary interactions were with four experienced strategy experts (\textbf{E1-E4}) from the team, as detailed in \autoref{tab:experts}. \textbf{E1}, the team manager, has eight years of experience in the esports industry, responsible for the overall operation and development of the team, including managing personnel changes, overseeing daily operations, providing psychological support, and fostering team building. \textbf{E2}, the coach, has transitioned from a professional player to a coaching role over four years, focusing on pre-match training, lineup selection, match strategies, and post-game analysis. \textbf{E3}, the team captain and Mid Laner, also serves as the shotcaller, making real-time strategic decisions and managing the team's playstyle, especially in the coach's absence. \textbf{E4}, the Top Laner, has collaborated closely with \textbf{E3} for three years, executing the established strategies in numerous matches.

\begin{table}[h]
    \begin{center}
    \caption{Profiles of participants in the observational study include details such as ID, gender, age, team role, and research experience.}
    \begin{tabular}{lccccc}
    \toprule
    \textbf{ID} & \textbf{Gender} &  \textbf{Age} & \textbf{Role}   & \textbf{Exp.} \\
    \hline
    E1     & Male & 29 & Team Manager & 8 \\
    E2     & Male & 28 & Team Coach & 4 \\
    E3     & Male & 22 & Professional Player (Shotcaller) & 4 \\
    E4     & Male & 21 & Professional Player & 3 \\
    \bottomrule
    \end{tabular}
    \label{tab:experts}
    \end{center}
\end{table}

\begin{table*}[h]\small
\centering
\caption{Detailed chronological record of players and coach behaviors and communications during post-game analysis using the traditional Tool -- the \textit{LOL Replay System}, with a detailed transcript of the most significant argument.}
\begin{tabular}{llll}
    \toprule
    \textbf{Time} & \textbf{Role} & \textbf{Behavior} & \textbf{Communication} \\ 
    \hline
    \parbox[t]{1.5cm}{\vspace{0mm}0:00-0:10\vspace{2mm}} & \parbox[t]{2.5cm}{\vspace{0mm} Coach \vspace{2mm}} & \parbox[t]{5cm}{\vspace{-1mm} Opened the \textit{LOL replay system} and launched the video of this match. \vspace{2mm}} & \\
    \hline
    \parbox[t]{1.5cm}{\vspace{0mm}0:10-5:00\vspace{2mm}} & \parbox[t]{2.5cm}{\vspace{0mm}5 Players\vspace{2mm}} & \parbox[t]{5cm}{\vspace{-1mm} Starting with the Shotcaller, followed by each player in turn, discussing and reflecting on their errors at specific moments. \vspace{2mm}} & \\
    \hline
    \parbox[t]{1.5cm}{\vspace{0mm}5:00-5:10\vspace{2mm}} & \parbox[t]{2.5cm}{\vspace{0mm} Coach \vspace{2mm}} & \parbox[t]{5cm}{\vspace{-1mm} \blueHigh{Directly pinpointed the major team fight}, \textbf{draging the progress bar to 30:30}. \vspace{2mm}} & \parbox[t]{5cm}{\vspace{-1mm} ``\textit{This drake fight is crucial. Who do you think is responsible for it?}'' \vspace{1mm}}\\
    \hline
    \parbox[t]{1.5cm}{\vspace{0mm}5:10-6:30\vspace{2mm}} & \parbox[t]{2.5cm}{\vspace{0mm} Shotcaller (Mid Laner) \vspace{2mm}} & \parbox[t]{5cm}{\vspace{-1mm} Focused the screen on the champions using the mouse. \vspace{2mm}} & \parbox[t]{5cm}{\vspace{-1mm} He first reflected on himself, ``\textit{I shouldn't have come over, it was useless. Should've just stayed in the river and blocked their carries instead of chasing `Jarvan'.}'' He continued, ``\textit{\greenHigh{I was gonna control the river, but `Jarvan' dropped an E+Q into the dragon pit out of nowhere.}}''\vspace{1mm}}\\
    \hline
    \parbox[t]{1.5cm}{\vspace{0mm} 6:30-15:40 \vspace{2mm}} & \parbox[t]{2.5cm}{\vspace{0mm} Bot Laner \vspace{2mm}} & \parbox[t]{5cm}{\vspace{-1mm} \textbf{Dragged the progress bar back to 30 seconds before the teamfight.} \vspace{2mm}} & \parbox[t]{5cm}{\vspace{-1mm} He questioned the Support's behavior,  ``\textit{\redHigh{`Jarvan' E+Q'd to place vision. Why didn't you seize the opportunity to engage on him?}}''\vspace{1mm}}\\
    \hline
    & \parbox[t]{2.5cm}{\vspace{0mm} Mid Laner \vspace{2mm}} & \parbox[t]{5cm}{\vspace{-1mm} \textbf{Repeatedly dragged the progress bar to observe the enemy ``Jarvan''.} \vspace{2mm}} & \parbox[t]{5cm}{\vspace{-1mm} He also expressed confusion, ``\textit{\redHigh{Looked crazy to me. `Jarvan' kept EQing right in our faces to clear wards.}}''\vspace{1mm}} \\
    \hline
    & \parbox[t]{2.5cm}{\vspace{0mm} Support \vspace{2mm}} & \parbox[t]{5cm}{\vspace{-1mm} \textbf{Dragged the progress bar to the teamfight} and pointed at the champion with the mouse. \vspace{2mm}} & \parbox[t]{5cm}{\vspace{-1mm} He explained, ``\textit{\redHigh{`Xin Zhao' has to E in to start the fight. `Rakan' can't engage first. I play `Rakan' as follow-up, using E to go in with the team, not to initiate.}}''\vspace{1mm}} \\
    \hline
    & \parbox[t]{2.5cm}{\vspace{0mm} Bot Laner \vspace{2mm}} & \parbox[t]{5cm}{\vspace{-1mm} \textbf{Dragged the progress bar back to before the teamfight again.} \vspace{2mm}} & \parbox[t]{5cm}{\vspace{-1mm} He further criticized the Support, ``\textit{\purpleHigh{Vision was key. `Jarvan' kept EQing for wards, and you didn't engage or set vision. That's why we lost the fight.}}''\vspace{1mm}} \\
    \hline
    
    \parbox[t]{1.5cm}{\vspace{0mm} 15:40-18:10 \vspace{2mm}} & \parbox[t]{2.5cm}{\vspace{-1mm} Bot Laner and Support \vspace{2mm}} & \parbox[t]{5cm}{\vspace{-1mm} Emotions ran high as both sides argued without reaching a conclusion. \vspace{2mm}} & \\
    \hline
    \parbox[t]{1.5cm}{\vspace{0mm} 18:10-23:20 \vspace{2mm}} & \parbox[t]{2.5cm}{\vspace{1mm} Coach \vspace{2mm}} & \parbox[t]{5cm}{\vspace{-1mm} Calmed the players' emotions and provided a conclusion. \vspace{2mm}} & \parbox[t]{5cm}{\vspace{-1mm} ``\textit{`Rakan' should've initiated when the Jungler couldn’t, and had several chances to pick `Jarvan' but missed.}''\vspace{1mm}}  \\
    \hline

    \parbox[t]{1.5cm}{\vspace{0mm} 23:20-70:30 \vspace{2mm}} & \parbox[t]{2.5cm}{\vspace{1mm} Coach \vspace{2mm}} & \parbox[t]{5cm}{\vspace{-1mm} \textbf{Rewound to the beginning}, starting the review anew, \blueHigh{only pausing at moments deemed critical for analysis}. Two more small arguments occurred among the five players. \vspace{2mm}} & \\
    \hline

    \parbox[t]{1.5cm}{\vspace{0mm} 70:30-73:50 \vspace{2mm}} & \parbox[t]{2.5cm}{\vspace{1mm} Support \vspace{2mm}} & \parbox[t]{5cm}{\vspace{-1mm} \blueHigh{After the whole video was reviewed, the Support noticed a significant mid-lane damage gap on the stats panel} and \textbf{rewound to the key teamfight.} \vspace{2mm}} & \parbox[t]{5cm}{\vspace{-1mm} He had a new take, ``\textit{\purpleHigh{How is it my fault? Their mid's poke was too strong. Our Top and Jungler got chunked. We lost because our carries were in poor condition, yet I had to follow the shotcall and force the engage.}}''\vspace{1mm}} \\
    \hline

    \parbox[t]{1.5cm}{\vspace{0mm} 73:50-80:10 \vspace{2mm}} & \parbox[t]{2.5cm}{\vspace{0mm} Support, Bot Laner and Coach \vspace{2mm}} & \parbox[t]{5cm}{\vspace{-1mm} Both sides argued again. The coach calmed them down and summarized each player's mistakes in the fight. \vspace{2mm}} & \parbox[t]{5cm}{\vspace{-1mm} ``\textit{First, poor vision due to the Jungler and Support. Second, lack of communication on who should initiate, and the shotcaller didn’t lead. Third, the Top and Jungler were out of position and got poked down, preventing from initiating. We should've capitalized on enemy mistakes, like picking `Jarvan' before the fight.}''\vspace{1mm}} \\
    \bottomrule
    
\end{tabular}
\label{tab:communications}
\end{table*}

\subsection{Contextual Inquiry of Post-game Analysis}
\label{sec:Contextual Inquiry}
\par To capture user workflows and authentic requirements, we conducted a contextual inquiry~\cite{raven1996using} by observing a recent post-game analysis with the collaborating team. We documented the behaviors and communications of all roles involved, as detailed in \autoref{tab:communications}. This led to the identification of key issues and limitations, resulting in five findings. Findings \prefix{\texttt{\textbf{[F1-F3]}}} highlight inefficiencies in traditional review tools, while \prefix{\texttt{\textbf{[F4-F5]}}} focus on human factors, such as subjective biases and entrenched habits that current tools fail to address.


\par \textbf{\prefix{\texttt{\textbf{[F1]}}} Inefficiency of Traditional Reviewing Process.} The conventional review process is time-consuming and inefficient. A typical 30-40 minute game requires about 80 minutes for a full review, and even excluding unnecessary disputes, it still takes around 60 minutes. A major issue is the operational challenges players face with traditional tools like the \textit{LOL Replay System}. They must manually adjust the video timeline to find critical moments (as shown in the \textbf{bolded section} of \autoref{tab:communications}), which is imprecise. Even experienced coaches often overshoot and need to fine-tune playback repeatedly, leading to frustration and prompting them to replay earlier segments. Additionally, reliance on personal judgment to assess strategic decisions necessitates reviewing non-essential segments to verify their relevance, further increasing time inefficiency.

\par \textbf{\prefix{\texttt{\textbf{[F2]}}} Absence of Evidence-Based Impact Assessment.} When using the official replay system to identify inconsistencies, players often struggle to quantify the impact and severity of these issues on subsequent game developments, relying heavily on personal experiences. For example, the Bot Laner attributed a lost teamfight and drake to the Support's failure to ``\textit{engage or set vision}'', while the Support claimed the loss resulted from the Top Laner and Jungler being poked down by the enemy Mid Laner (as highlighted in \purpleHigh{purple} in \autoref{tab:communications}). This unresolved disagreement illustrates the current review process's inability to measure the impact of inconsistent behaviors on outcomes, limiting the depth and quality of insights derived from the analysis.

\par \textbf{\prefix{\texttt{\textbf{[F3]}}} Lack of Understanding of Opponent's Playstyle.} During both the execution and evaluation of a match, team members often struggle with fully grasping the opponent's unique playstyle. This issue is exemplified in \autoref{tab:communications}, as highlighted in \greenHigh{green}, where the Jungler initially ``\textit{planned to control the river}'' but was unaware of the opponent's aggressive tendencies, leading to an unexpected ``\textit{E+Q into the dragon pit}''. A holistic understanding of the opponent's strategies, particularly in the pre- and mid-game phases, is essential for crafting effective tactics. Moreover, this awareness is vital in post-match reviews for pinpointing the root causes of strategic inconsistencies. These insights are invaluable for team leaders looking to refine and enhance overall decision-making.

\par \textbf{\prefix{\texttt{\textbf{[F4]}}} Assumptions and Subjective Biases.} Differing interpretations of in-game situations often lead to strategy discussions influenced by personal experiences and biases, as noted in \redHigh{red} in \autoref{tab:communications}. Players frequently make assumptions; for example, the Bot and Mid Laner criticized the Support for indecisiveness in engaging, while the Support maintained that his role was to follow the Jungler's engage or counter-initiate. This illustrates the current post-game analysis process's susceptibility to subjective interpretations. Without objective data, reaching consensus on preferred professional strategies becomes challenging, hindering the understanding of in-game collaboration and actual inconsistencies. Consequently, this traditional method can shift focus to disputes and self-justification, undermining team cohesion and potentially impacting future teamwork negatively.

\par \textbf{\prefix{\texttt{\textbf{[F5]}}} Prioritizing Major Team Fights Over Crucial Details.} In post-game analysis, teams often emphasize key battles that significantly impact the game's outcome, as noted in \blueHigh{blue} in \autoref{tab:communications}. \textbf{E3} explained, ``\textit{The most critical team fight is usually a consensus, with rarely any disagreements,}'' citing efficiency as the main reason: ``\textit{Analyzing multiple team fights takes so long that detailing every small skirmish would take hours.}'' While focusing on major fights, players often overlook other crucial moments that collectively influence the game's result. For instance, only the Support acknowledged the poke difference between Mid Laners late in the review, highlighting missed opportunities to improve. This lack of thorough examination of each key moment can hinder the team's overall development.


\subsection{Key Considerations and Design Requirements}
\par Our exploration involved two 60-minute phases of semi-structured interviews with \textbf{E1-E4}. The first phase focused on the importance of examining inconsistencies between preferred professional and real-time strategies, identifying critical factors that influence strategic decision-making and their hierarchical significance (\textbf{RQ1}). We gathered detailed background information on the esports team, including its history, duration, role distribution, and overall size, along with their competitive history in major tournaments and achievements. Additionally, we identified and prioritized key factors affecting decision-making in professional gaming teams, summarized below.
\begin{itemize}
    \item \prefix{\texttt{\textbf{[KC1]}}} \textbf{Individual Status of Both Allies and Enemies.} Expert consensus highlights this factor as critical, which includes health points (hp) and mana points (mp), is considered the most critical, along with economic status, equipment, and level. \textbf{E2} illustrated this by stating, ``\textit{If Player A has advantages over B in economy, equipment, and level, but is low on HP or out of mana while B is at full capacity, A should avoid engaging easily, as caution is essential in professional matches.}'' Additionally, \textbf{E3} stressed the importance of assessing both allies' support capabilities and the potential threats posed by opponents. He provided a straightforward example: ``\textit{If our Top Laner is in a better state than the enemy, yet the enemy Jungler and Mid Laner are unaccounted for, and our other four teammates are ganking the bot lane, our Top Laner should not advance recklessly, as he may become vulnerable to a gank without support.}''
    
    \item \prefix{\texttt{\textbf{[KC2]}}} \textbf{Resource Accumulation of Both Allies and Enemies.} Resource accumulation includes economic status, equipment, and level. As \textbf{E3} noted, ``\textit{Higher levels unlock or enhance skills, while greater economic resources allow for earlier acquisition of key equipment, which boosts a character's combat effectiveness.}'' Furthermore, \textbf{E3} highlighted that increased resource accumulation enhances map control, providing greater vision and mobility, which in turn facilitates the suppression of opponents.
    
    \item \prefix{\texttt{\textbf{[KC3]}}} \textbf{Positions of Other Allies and Enemies.} Strategic decision-making relies heavily on evaluating the positions of both allies and enemies. \textbf{E3} emphasized that ``\textit{If the enemy Jungler is ganking the bot lane, our Jungler, if nearby, can quickly secure the Rift Herald.}'' Furthermore, \textbf{E3} underscored the significance of monitoring the ``Respawn Times of Key Resource Points'', such as red and blue buffs, Rift Herald, Drakes, and Baron Nashor. However, \textbf{E2} provided a nuanced perspective, ``\textit{While the respawn times of key resources are undoubtedly important, they are often indicated by player positioning. For instance, as a drake approaches its respawn, players naturally converge in the bot lane.}'' This highlights the intricate interplay between player positioning and the impending respawn of critical in-game resources.
\end{itemize}

\par In the second phase of our interviews, we consulted with experts to gain insights into the limitations and challenges of traditional methods and tools, as well as to understand their primary concerns and requirements. Utilizing thematic analysis of the gathered feedback~\cite{terry2017thematic}, we identified key insights that informed the establishment of design requirements (\prefix{\texttt{\textbf{[DR1]}}} - \prefix{\texttt{\textbf{[DR4]}}}) for our potential approach.


\par \textbf{\prefix{\texttt{\textbf{[DR1]}}} Thorough Identification of Inconsistencies.} To comprehensively uncover and swiftly pinpoint all points of inconsistency, we revisited the entire post-game analysis workflow. \textbf{E2} explained, ``\textit{Like our debriefs, we usually start by reviewing the most impactful team fight. Then, either I or the team captain guides the discussion through the match, pointing out areas we feel were executed poorly \prefix{\texttt{\textbf{[F5]}}}}.'' \textbf{E4} emphasized the time-consuming nature of this traditional method, noting, ``\textit{Debriefing a game that lasts 30-40 minutes often takes an hour or more, and if there are disagreements, it can take even longer \prefix{\texttt{\textbf{[F1]}}}}.'' The reliance on experience and intuition, along with visual observation and frequent pauses in the footage, limits the effectiveness of this approach in identifying inconsistencies comprehensively. Moreover, the sequential review of the entire game footage adds a significant time burden, making the process both thorough and inefficient. Recognizing these challenges, our approach should provide a methodology that is not only comprehensive but also efficient in revealing strategic inconsistencies.

\par \textbf{\prefix{\texttt{\textbf{[DR2]}}} Quantifying the Impact of Inconsistencies on Subsequent Gameplay.} During our consultation with coaches on how they assess the significance of one team fight relative to another, \textbf{E2} explained, ``\textit{It's usually a collective decision. If most of us believe a specific fight caused the loss, we start the debriefing from there.}'' However, disagreements are common, often influenced by subtle psychological factors. \textbf{E2} provided an example, ``\textit{If A made a significant inconsistency in the first fight and B made one in the second, A might view B's inconsistency in the second fight as the critical moment that caused the loss, while B may point to A's earlier inconsistency. It's difficult to quantify these moments, like losing a drake in one fight versus a key tower in another \prefix{\texttt{\textbf{[F2]}}}.}'' 
This underscores the need for a method to accurately quantify the impact of inconsistencies on subsequent gameplay. Such a method would not only clarify accountability and promote team harmony but also enhance the team's understanding of the long-term effects of different tactical decisions.

\par \textbf{\prefix{\texttt{\textbf{[DR3]}}} Facilitating Rapid Understanding of Teams' Playstyles and Historical Performances.} \textbf{E1} highlighted the resource constraints faced by smaller esports clubs, noting, ``\textit{Many smaller esports clubs can't afford a dedicated data analyst, so the coach and I handle data analysis together.}'' \textbf{E2}, discussing the challenges of studying other teams' playstyles, added, ``\textit{Before matches, we review our opponents' past games, usually recording their historical performance in Excel with basic metrics like average damage and economy.}'' However, as emphasized by \textbf{E2}, this method only captures fundamental data, offering limited insights \prefix{\texttt{\textbf{[F3]}}}. He further remarked, ``\textit{Pure numbers are dry and abstract, lacking interaction.}'' To address these limitations, we propose offering a more comprehensive perspective that enables players to swiftly grasp both teams' playstyles, historical performances, and individual preferences. This enhancement is designed to offer deeper insights into inconsistent behaviors, facilitating more informed and strategic planning.
\begin{figure*}[h]
  \centering
  \includegraphics[width=\textwidth]{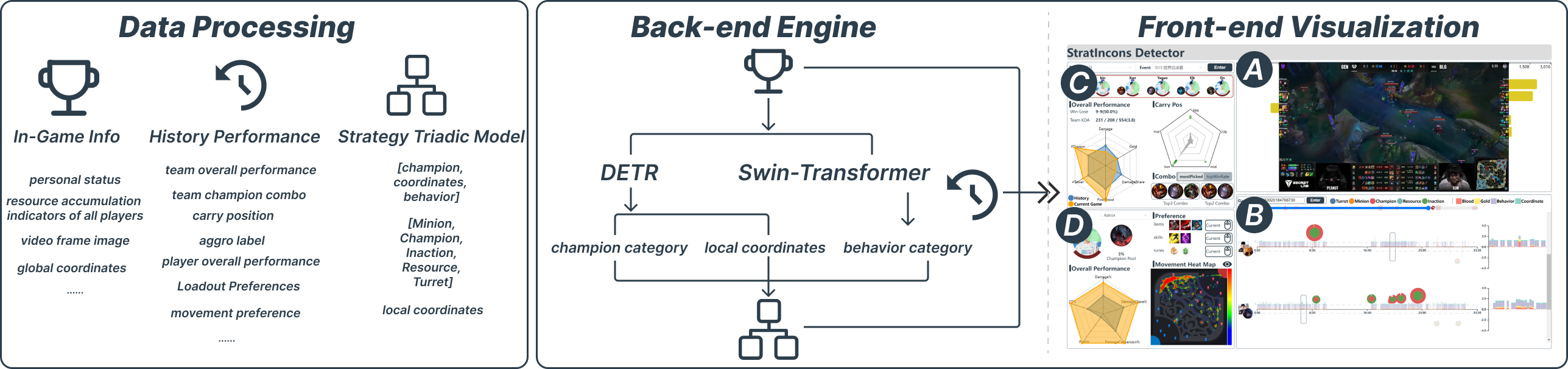}
  \caption{The \textit{StratIncon Detector} pipeline includes: \textit{Data Processing} for collecting and processing data, \textit{Back-end Engine} for model training and strategy prediction, and \textit{Front-end Visualization} for detecting inconsistencies and providing match insights.}
  \label{fig:system_pipeline}
\end{figure*}
\par \textbf{\prefix{\texttt{\textbf{[DR4]}}} Assist in Identifying Preferred Professional Strategies.} According to \textbf{E3} (Team Captain), conflicts frequently arise during routine training matches when the coach is absent \prefix{\texttt{\textbf{[F4]}}}. He stated, ``\textit{I usually lead the debriefs, but conflicts come up because everyone has different opinions. We really need an impartial third party to help us figure out the best strategy in those situations.}'' In response, our approach should employ model algorithms to assess inconsistencies in actions and suggest a potential definitive preferred professional strategy, moving beyond players' subjective interpretations and experiences with champions and situations. Given the uncertainties inherent in MOBA games, professional players must consider a range of possible strategies; therefore, the model should present the top $n$ strategies. This approach aims to minimize disputes stemming from subjective biases, enhance the efficiency of team debriefings, and promote team cohesion.

\section{StratIncon Detector}
\par Based on insights from our observational study with experts, we developed the interactive visual analytics system \textit{StratIncon Detector} to help professional teams identify strategic inconsistencies and evaluate their impact on gameplay, as illustrated in \autoref{fig:system_pipeline}. In the \textbf{Data Processing Phase}, we collect and process \textit{Real-Time In-Game Information} and \textit{Historical Match Performance} data, integrating expert insights. During the \textbf{Back-end Engine Phase}, we train a \textit{DETR} model on manually annotated images to detect champion categories and local coordinates, while a \textit{Swin-Transformer} recognizes behavioral categories. This culminates in a comprehensive strategy triadic model for each frame across all match data. Additionally, we train an \textit{LSTM} model to predict the strategy triadic model for the next frame based on the previous five frames. In the \textbf{Front-end Visualization Phase}, the system offers the following functionalities: (1) Predictions of the preferred professional strategy at any moment \textbf{\prefix{\texttt{\textbf{[DR4]}}}}; (2) Quick identification of strategic inconsistencies for all players \textbf{\prefix{\texttt{\textbf{[DR1]}}}}; (3) Quantitative analysis of the impact of these inconsistencies on gameplay \textbf{\prefix{\texttt{\textbf{[DR2]}}}}; and (4) Insights into team and player playstyles and historical preferences to trace the origins of these inconsistencies \textbf{\prefix{\texttt{\textbf{[DR3]}}}}.

\subsection{Data Processing}
\par The dataset for this study is sourced from the official Riot esports tournament platforms, including \textit{LOL Esports}\footnote{https://lolesports.com}, \textit{scoregg}\footnote{https://www.scoregg.com/}, and \textit{escorenews}\footnote{https://escorenews.com/en}. It encompasses a range of international events from 2023, such as major tournaments like \textit{Worlds}, \textit{MSI}, and \textit{LCS}, totaling 50 professional matches. We collaborated with experts (\textbf{E1-E4}) to manually verify these matches, ensuring no critical errors impacted outcomes. The dataset includes both match videos and in-game logs for teams and players, along with historical performance data from \textit{scoregg}. For a quantitative analysis of strategic inconsistencies, we synthesized key considerations and findings, processed the metadata, and organized the data into the following subdivisions.

\par \textbf{Real-Time In-Game Information:} \textit{1) Player Status \textbf{\prefix{\texttt{\textbf{[KC1]}}}}}: This includes attributes for each player at the current frame, including \textit{HP}, \textit{mana}, \textit{physical attack}, \textit{magical attack}, \textit{movement speed}, and \textit{attack speed}. \textit{2) Resource Accumulation Indicators \textbf{\prefix{\texttt{\textbf{[KC2]}}}}}: This refers to the economic and level data for each player in the current frame. \textit{3) Video Frame Image \textbf{\prefix{\texttt{\textbf{[KC3]}}}}}: The broadcaster's perspective\footnote{In MOBA esports matches, the \textit{Broadcaster's perspective} refers to the view provided by the broadcast team, highlighting key gameplay elements and player actions to enhance the viewing experience.} of the current frame, capturing all environmental elements within the game. \textit{4) Global Coordinates \textbf{\prefix{\texttt{\textbf{[KC3]}}}}}: The mini-map located in the bottom right corner of the screen, providing real-time global positioning data for all $10$ players.

\par \textbf{Historical Match Performance:} \textit{1) Team Overall Performance:} Metrics such as win rate, team KDA, damage, gold earned, damage taken, and first blood rate in a specific event. \textit{2) Team Champion Combo:} The most frequently used champion pairings and those with the highest win rates. \textit{3) Carry Position Preference:} Indicates which position is prioritized as the team's carry, receiving most resources. \textit{4) Aggro Label:} The team's general playstyle, whether it leans towards a more conservative or aggressive approach. \textit{5) Player Overall Performance:} Statistics like KDA, champion pool, damage contribution, damage taken, damage conversion, team fight participation, and last-hit accuracy. \textit{6) Loadout Preferences:} The player's favored items, skills, and runes when playing a specific champion. \textit{7) Movement Patterns:} The player's common positions on the map when using a particular champion.

\begin{figure*}[ht]
  \centering
  \includegraphics[width=\textwidth]{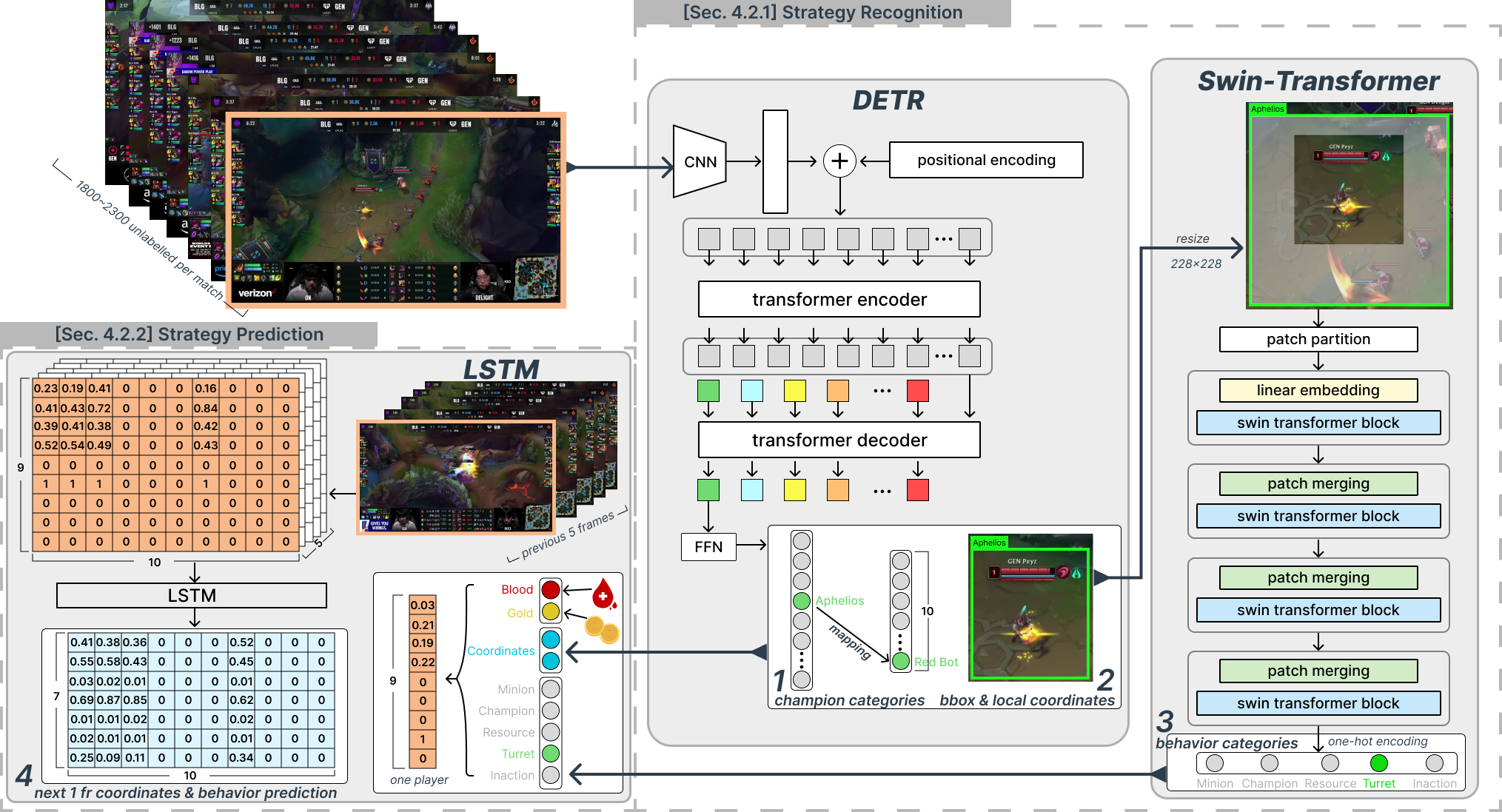}
  \caption{\textit{StratIncon Detector} Back-end details: During strategy recognition, unlabelled match frames are processed by \textit{DETR} to extract (1) champion categories and (2) coordinates. Resized champion slices are classified into (3) behavior categories using the \textit{Swin-Transformer}. In the prediction phase, the previous five frames are input into \textit{LSTM} to predict (4) the next frame's coordinates and behavior. Each player's \textit{Blood, Gold}, \textit{Coordinates}, and \textit{Behavior} are encoded into a 9-dimensional vector per frame.}
  \label{fig:backend}
\end{figure*}
\par \textbf{Strategy Triadic Model:} The concept of strategy, though abstract, is inspired by tactics and strategies in team sports~\cite{wu2018forvizor,tian2019use}. In basketball, tactics encompass player positions (coordinates), roles/actions (behaviors), and player interactions. Similarly, MOBA games focus on skills, positions, and behaviors to improve team performance~\cite{tian2019use}. We define strategy as a triad of [champion, coordinates, behavior], representing the ``who'', ``where'', and ``what'' in gameplay, forming the basis of the strategy triadic model. Expert feedback, particularly from \textbf{E2}, highlighted the difficulty of precisely defining strategy in MOBA games, though certain abstract definitions can achieve consensus. Strategic behaviors are categorized into five groups: [Minion, Champion, Resource, Turret, Inaction]. For instance, if a champion's behavior is classified as ``Minion'', it indicates the champion is attacking minions in that frame. Further details on all behaviors are provided in \autoref{appendix:behaviors}.

\par Coordinates in MOBA games involve both local coordinates (pixel coordinates relative to the top-left corner of the current video frame) and global coordinates (absolute global coordinates on the minimap, as illustrated in the \autoref{fig:Player_Preference_View}(D2)). Experts emphasized the significance of both types, noting that local coordinates provide more scene-specific information, especially for short-term behavior prediction.


\begin{figure}[ht]
    \centering
    \includegraphics[width=\linewidth]{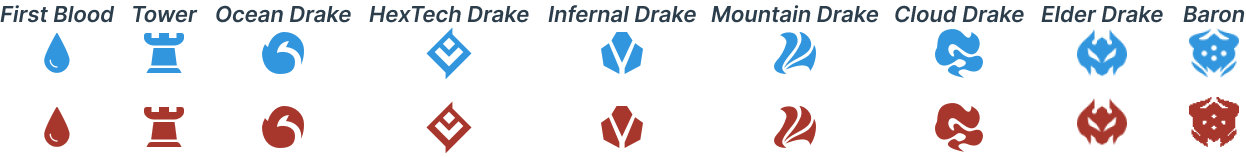}
    \caption{All event icons for blue and red teams.}
    \label{fig:events}
\end{figure}

\par \textbf{Global Priority Events:} Similar to previous research on MOBA games~\cite{jin2023actorlens,yang2022predicting}, this study also defines key events (\autoref{fig:events}), but approaches them from the perspective of professional esports. During the observational study interviews, \textbf{E2} explained, ``\textit{In professional matches, there are way fewer kills than in amateur games, which might not seem like a big deal, but it's actually really important. Pros understand that the game is about pushing towers, not just getting kills.A kill doesn’t win the game—it’s what happens after, like losing a tower or missing a drake fight, that matters.}'' \textbf{E3} further added, ``\textit{In our regular training, towers and epic monsters are the global events we focus on the most.}''

\par Experts highlighted the importance of prioritizing global priority events such as tower pushes and epic monsters contests over more granular incidents like kills or buffs. However, our study does not overlook these detailed events. Instead, they are incorporated into our analysis, as they often contribute to broader level and economic trends~\cite{li2016visual}, which are captured as part of the ``Real-Time In-Game Information''.

\par \textbf{Inconsistency Impact:} In games with win or lose outcomes, it is possible to assess the marginal impact of primary actions on the final result~\cite{christiansen2021deployment}. However, this study focuses on the impact of inconsistencies on each subsequent event. \textbf{E2} mentioned that the outcomes of nearly all events are directly reflected in the economic curve, ``\textit{Destroying towers, securing kills, and capturing neutral resources boost economy, leading to better gear and higher chances of winning}''. Therefore, using the normalized economic difference between the moment an inconsistency occurs and the subsequent event to quantify the impact of the inconsistency is a method that is both interpretable and accepted by experts.

\begin{table*}[ht]
  \centering
  \caption{Performance of \textbf{DETR}, \textbf{Swin-Transformer} and \textbf{LSTM}.}
  \begin{tabular}{lcccccccc}
    \toprule
    \textbf{Task}& \textbf{Model} & \textbf{Train} & \textbf{Validation} & \textbf{Test} & \textbf{F1(\%)} & \textbf{ACC(\%)} & \textbf{MSE} & \textbf{MAE} \\
    \midrule
    Champion \& Coordinate Recognition& \textbf{DETR} & \textbf{7523} & \textbf{462} & \textbf{2015} & \textbf{88.75} & \textbf{89.93}  & - & - \\
    &YoloV5 & 7523 & 462 & 2015 & 82.53 & 79.88 & - & - \\
    \midrule 
    Behavior & \textbf{Swin-Transformer} & \textbf{7523} & \textbf{462} & \textbf{2015} & \textbf{94.19} & \textbf{93.34} & - & - \\
    &ResNet18 & 7523 & 462 & 2015 & 92.04 & 90.87 & - & - \\
    \midrule 
    Prediction & \textbf{LSTM} & \textbf{7523} & \textbf{462} & \textbf{2015} & - & - & \textbf{0.303} & \textbf{0.312} \\ 
    &Transformer & 7523 & 462 & 2015 & - & - & 0.331 & 0.342  \\
    \bottomrule
  \end{tabular}
  \label{tab:backend_res}
\end{table*}

\subsection{Back-end Engine}
\par The backend of our system comprises two main components: \textit{strategy recognition} and \textit{strategy prediction}. For strategy recognition, we use \textit{DETR} to classify champion categories and extract local coordinates, while the \textit{Swin-Transformer} identifies behavior categories. This dataset is then processed by an \textit{LSTM} model, trained to predict the strategy triadic model for future frames based on historical patterns. Performance metrics for \textit{DETR}, \textit{Swin-Transformer}, and \textit{LSTM} are detailed in \autoref{tab:backend_res}, with experimental specifics in \autoref{tab:backend_para}.

\subsubsection{Strategy Recognition}
\par The temporal prediction model used in this study, specifically the LSTM (\autoref{label:strategy_prediction}), relies on extensive multi-dimensional input data, including \textit{champion categories}, \textit{local coordinates}, and \textit{behavior categories}. To obtain these data points, which are not directly accessible, we utilized the \textit{DETR} and \textit{Swin-Transformer} models to extract and impute the necessary information across all frame images in the dataset.

\par \textbf{Champion Categories \& Local Coordinates.} To determine champion categories and local coordinates (relative to the top-left corner of the current video frame), we employed the \textit{DETR}~\cite{carion2020end}. This model is highly effective for object detection, ensuring precise localization and classification of champions. We annotated champion categories and local coordinates in 200 images per match using \textit{labelme}\footnote{https://github.com/labelmeai/labelme}, following the \textit{COCO (Common Objects in Context)} dataset format~\cite{lin2014microsoft}. Subsequently, \textit{DETR} was applied to recognize the remaining frame images for each match. As shown in \autoref{fig:backend}, the model processes a three-channel input image, producing the local coordinates of all champions in that frame along with their corresponding 10-dimensional one-hot vectors.

\par \textbf{Behavior Categories.} After identifying champion categories and local coordinates, the next step is to determine the corresponding behavior categories. For this task, we leveraged the \textit{Swin-Transformer}~\cite{liu2021swin}, which is renowned for its exceptional performance on small sample datasets. To capture additional contextual information, we slightly expanded the champion location area obtained from \textit{DETR} to a size of (228, 228). As shown in \autoref{fig:backend}, this adjusted image is then input into the \textit{Swin-Transformer}, which generates a 5-dimensional one-hot vector representing the behavior categories.

\par Consequently, we obtained a strategy triadic model for each champion in the frame, yielding approximately $2,000$ to $2,500$ frame images per match.

\subsubsection{Strategy Prediction}
\label{label:strategy_prediction}

\begin{table}[h]
    \begin{center}
        \caption{The experiment details (Input Shape, Epoch, Batch Size, Initial Learning Rate, Device) of \textbf{DETR}, \textbf{Swin-Transformer} and \textbf{LSTM}.}
        \scalebox{0.82}{ 
        \begin{tabular}{lccccc}
        \toprule
            \textbf{Model} & \textbf{InputShape} & \textbf{Epoch} & \textbf{BatchSize} & \textbf{LR} & \textbf{Device}\\
        \midrule
             \textbf{DETR} & (1920,1080,3)&300& 16 & 2e-3 & RTX3090 \\
             \textbf{Swin-Transformer} & (228,228,3)&300& 32 & 1e-4 & RTX3090 \\
             \textbf{LSTM} & (9,10,5)&1000& 16 & 3e-4 & RTX3090 \\
        \bottomrule
        \end{tabular}
        }
        \label{tab:backend_para}    
    \end{center}
    
\end{table}

\par Following the strategic recognition, we employed an \textit{LSTM} model~\cite{hochreiter1997long} to predict future frames in the replay video. This model uses information from the previous five frames to forecast the next one. To streamline the input data and mitigate the influence of less relevant factors, we processed the input for each champion as shown in \autoref{fig:backend}. The coordinates indicate the champion's current location, derived from the \textit{DETR} output. Behavior is represented as a 5-dimensional one-hot vector based on the \textit{Swin-Transformer} output. Additionally, champion data includes the champion's current blood (HP) and gold value, normalized into a two-dimensional vector. Consequently, each champion's information is encapsulated in a 9-dimensional vector. With 10 champions per match and an input spanning five frames, the final input for the \textit{LSTM} model is a vector of dimensions 9*10*5. The \textit{LSTM} output frame data consists of [Coordinates, Behavior], resulting in a vector of dimensions 7*10.

\subsection{Front-end Visualization}

\begin{figure*}[h]
  \centering
  \includegraphics[width=\textwidth]{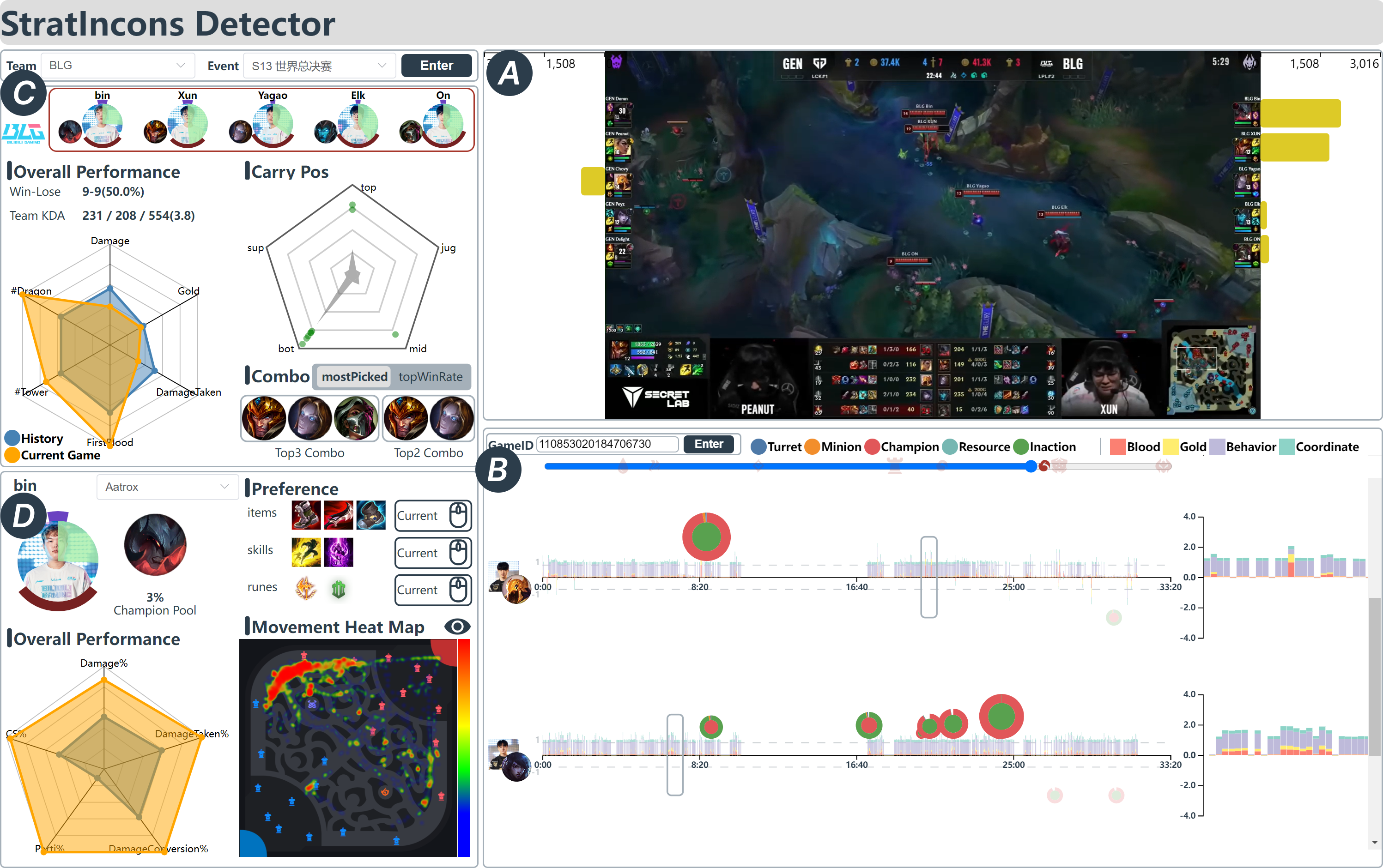}
  \caption{\textit{StratIncon Detector} Front-end Overview: (A) Replay View. (B) Player Inconsistency View. (C) Team Performance View. (D) Player Preference View.}
  \label{fig:overview}
\end{figure*}

\par Guided by the visualization design mantra of ``overview first, zoom and filter, then details-on-demand''~\cite{shneiderman1996eyes}, we developed the \textit{StratIncon Detector} (\autoref{fig:overview}). Iteratively refined through expert feedback, the system includes four views: the \textit{Replay View}, the \textit{Player Inconsistency View}, the \textit{Team Performance View}, and the \textit{Player Preference View}. Users start with the \textit{Replay View} and \textit{Player Inconsistency View} to review key events, analyze strategy inconsistencies, and their distribution among players (\prefix{\texttt{\textbf{[DR4]}}}). These views link inconsistencies with video footage and subsequent events (\prefix{\texttt{\textbf{[DR1]}}}), enabling quantitative evaluation of their impact on the game (\prefix{\texttt{\textbf{[DR2]}}}). The \textit{Team Performance View} and \textit{Player Preference View} then provide historical data on team and player performance, including insights into player preferences, to help users better understand the causes of these inconsistencies (\prefix{\texttt{\textbf{[DR3]}}}).

\subsubsection{Replay View}

\par
\begin{figure}[h]
    \includegraphics[width=\linewidth]{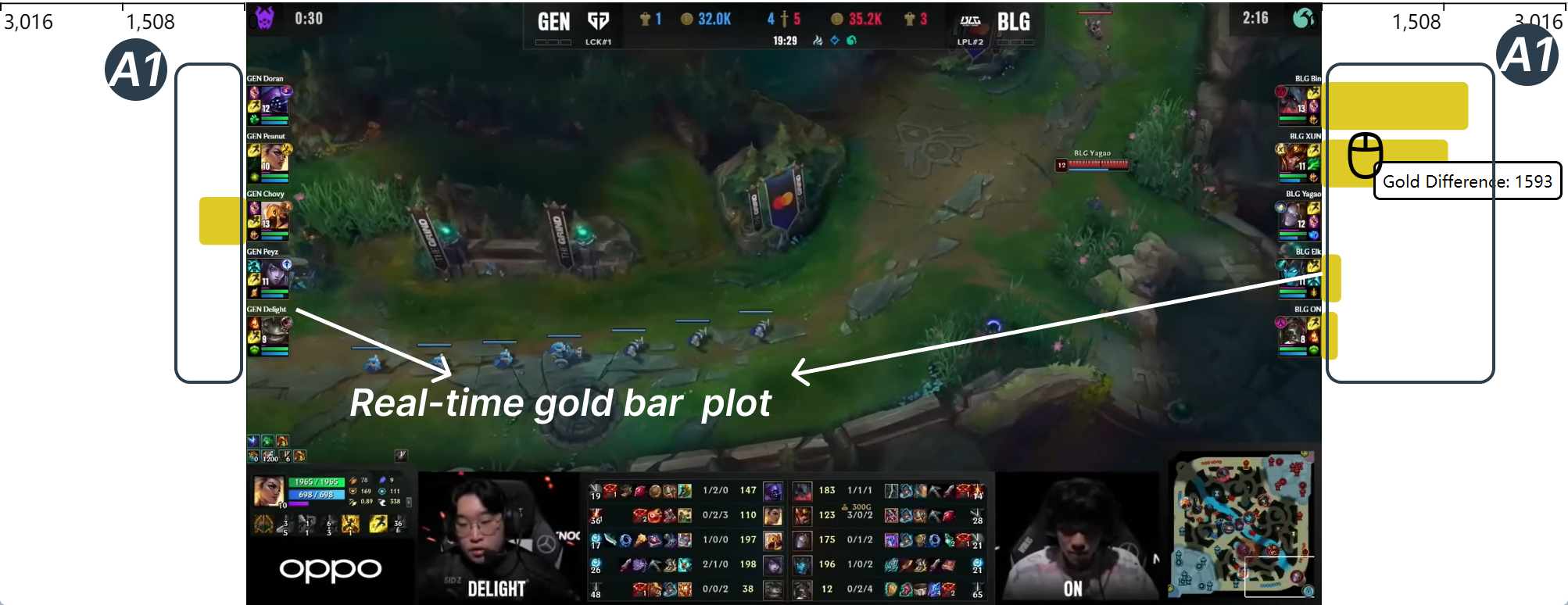}
    \caption{Overview of Replay View: (A1) Real-Time Gold Bar Plot.}
    \label{fig:replay_view}
\end{figure}

 To enhance user understanding while minimizing the learning curve and accurately capturing player behavior, we seamlessly integrated original game footage into the system. This allows a real-time bar chart (\autoref{fig:replay_view}) to overlay the video, visually illustrating economic disparities between players on opposing teams. Users can hover over the gold bar to view specific economic differences. Additionally, discrepancies in local coordinates are highlighted within the video frame, showing the real-time positions and their projected preferred professional counterparts. This approach facilitates empirical analysis of player strategies and actions. The \textit{Replay View} further enables users to retrospectively compare and review in-game behaviors.

\subsubsection{Player Inconsistency View}
\par The \textit{Player Inconsistency View} features two sub-views, \textbf{Event Timeline} and \textbf{Inconsistency Detector}. Together, they display all inconsistencies made by the losing players, offering users a platform to explore the root causes and assess their impact.

\par
\begin{figure*}[h]
  \centering
  \includegraphics[width=\textwidth]{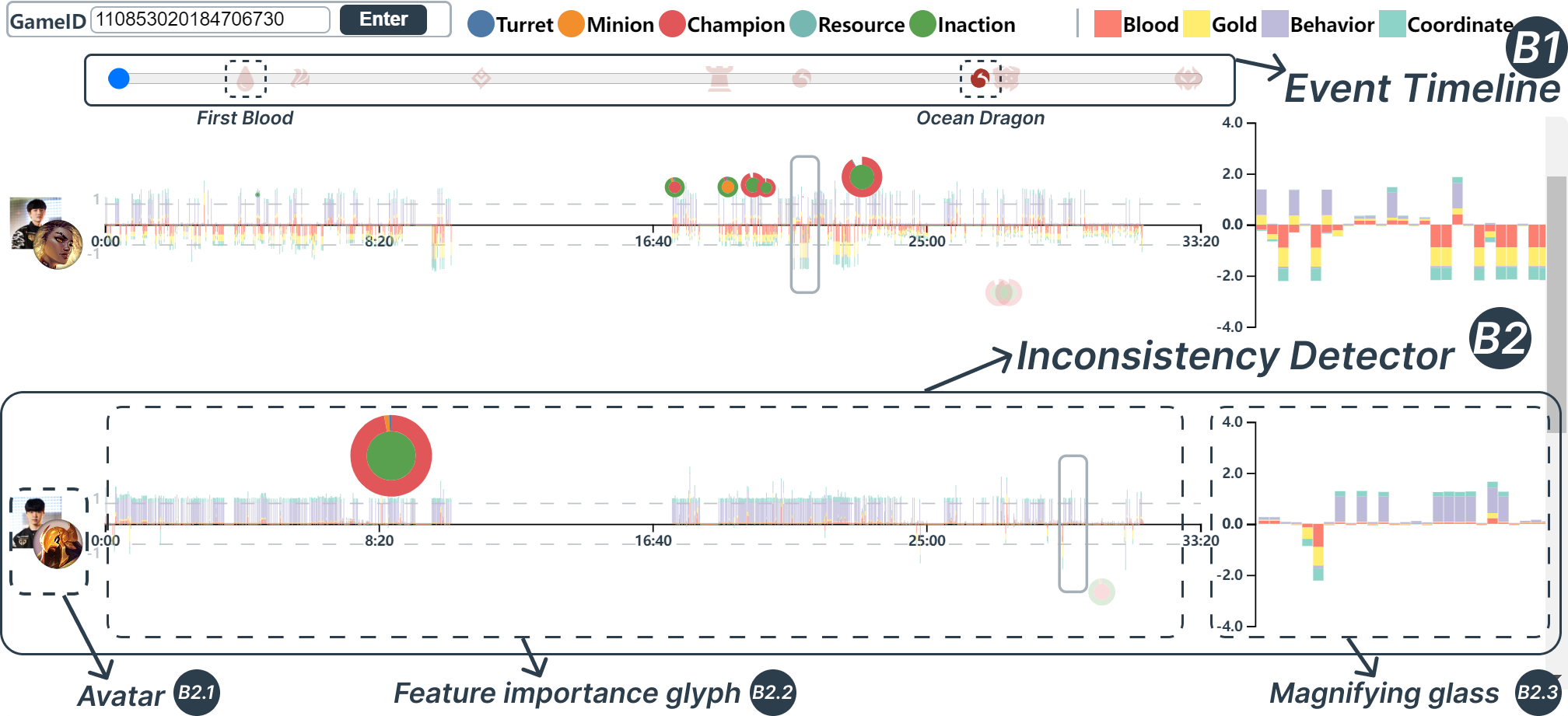}
  \caption{Overview of Player Inconsistency View: (B1) Event Timeline. (B2) Inconsistency Detector: (B2.1) Avatar. (B2.2) Feature Importance Glyph. (B2.3) Magnifying Glass.}
  \label{fig:Player_Inconsistency_View}
\end{figure*}

 \textbf{Event Timeline.} Positioned at the top of the view and synchronized with the source video, the timeline (\autoref{fig:Player_Inconsistency_View}(B1)) allows users to focus on key in-game events by dragging the progress bar. Icons (\autoref{fig:events}) represent major events like First Blood, First Tower, Drakes kills, and Baron Nashor kills, with colors indicating team affiliation. For example, a red blood event icon in \autoref{fig:Player_Inconsistency_View}(B1) signifies that the red team secured the First Blood, while a red ocean event icon indicates that the red team killed the Ocean Drake. Conversely, a red tower icon indicates that the first tower destroyed was the blue team's, highlighting that it was taken down by the red team. The currently highlighted icon signifies the specific event under focus.

\par \textbf{Inconsistency Detector.} Below the event timeline, five inconsistency detectors are arranged in the sequence of ``Top, Jungler, Mid, Bot, Support'' (\autoref{fig:Player_Inconsistency_View}(B2)), matching the commonly used order in the game and competitions. Each inconsistency detector consists of three main sections: left, center, and right. On the left side, the avatar section (\autoref{fig:Player_Inconsistency_View}(B2.1))  displays the avatars of the losing players along with their champion lineup for the match. In the middle section, there are the feature importance glyphs and inconsistency glyphs. The timeline in this section corresponds to the total duration of the video, with each second represented by a feature importance glyph. These glyphs evaluate the contribution of four features to predicting the preferred professional strategy at any given moment. As shown in  \autoref{fig:Player_Inconsistency_View}(B2.2), each feature importance glyph is a stacked bar plot where different colors represent various features, and the height of each bar encodes the normalized importance value. Features above the line have a positive impact on the prediction outcome, while those below the line have a negative impact. Users can use a brush box, linked to the magnifying glass section (\autoref{fig:Player_Inconsistency_View}(B2.3)) on the right, to focus on specific time periods of the feature importance bar plot. Design alternatives and additional details for the Inconsistency Detector are presented in \autoref{appendix:alt_inconsDetector}.


\par The inconsistency glyph appears only when there is a discrepancy between real-time and preferred professional strategies. It features an inner circle and an outer ring. The inner circle represents behaviors related to the real-time strategy, while the outer ring displays the top three potential preferred professional strategies as suggested by the model. The arc of each segment corresponds to the probability of each behavior. Different colors encode distinct behaviors, and the glyph's position aligns with the time of the inconsistency. The radius of the glyph indicates its impact on the selected event in the event timeline, with a larger radius signifying a more significant impact. This detailed arrangement allows for a dynamic analysis of player strategy inconsistencies throughout the match. Additionally, the glyph incorporates various interactive elements. Users can click it to jump to the corresponding time point in the original video within the \textit{Replay View}, enabling a direct and detailed comparison with the real video. Furthermore, selecting a different event icon on the timeline dynamically updates the glyph sizes to show their impact on the new event.

\subsubsection{Team Performance View}
\par
\begin{figure*}[ht]
    \begin{minipage}[t]{0.47\textwidth}
        \centering
        \includegraphics[width=\linewidth]{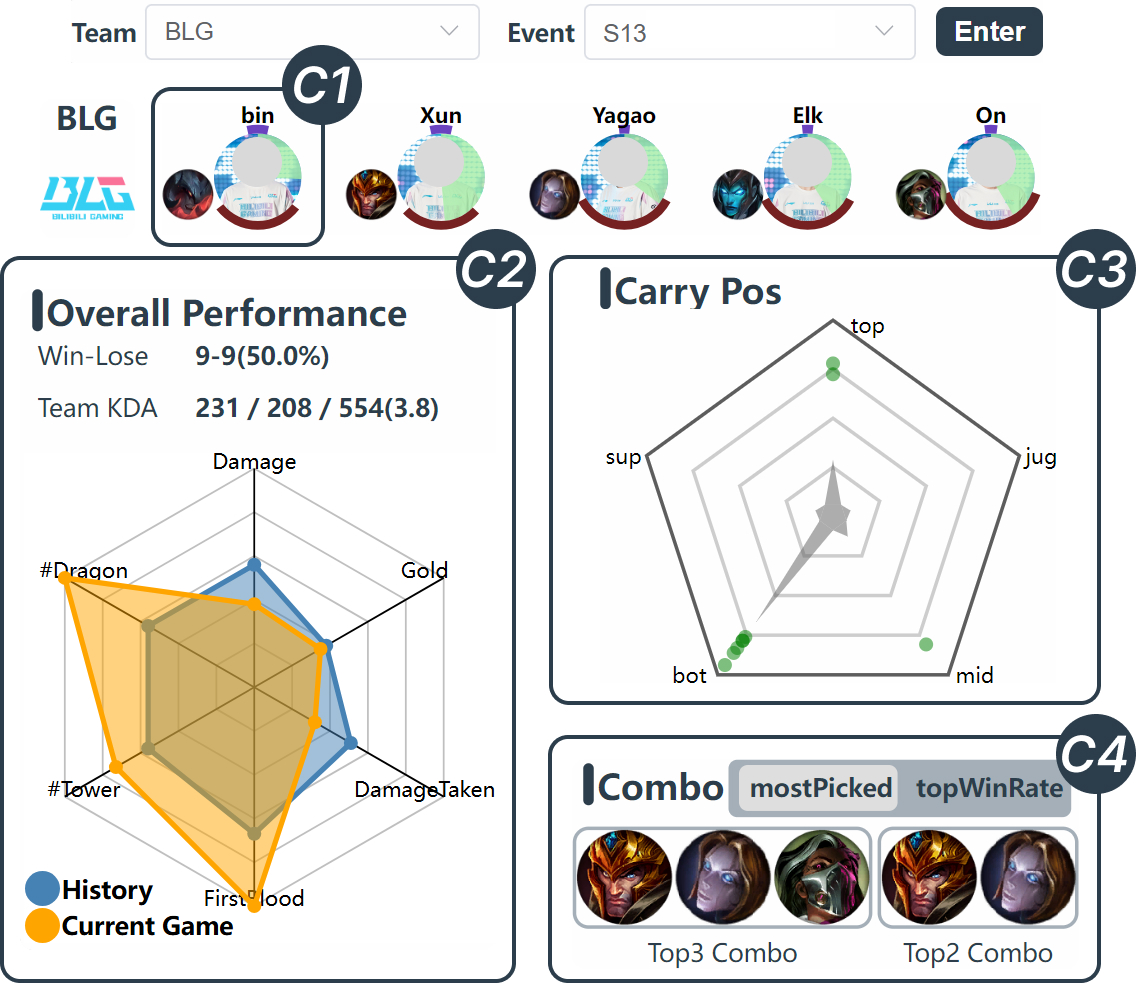}
        \caption{Overview of Team Performance View: (C1) Player Glyph. (C2) Overall Performance. (C3) Carry Position. (C4) Lineup Combo.}
        \label{fig:Team_Performance_View}
    \end{minipage}%
    \hfill
    \begin{minipage}[t]{0.47\textwidth}
        \centering
        \includegraphics[width=\linewidth]{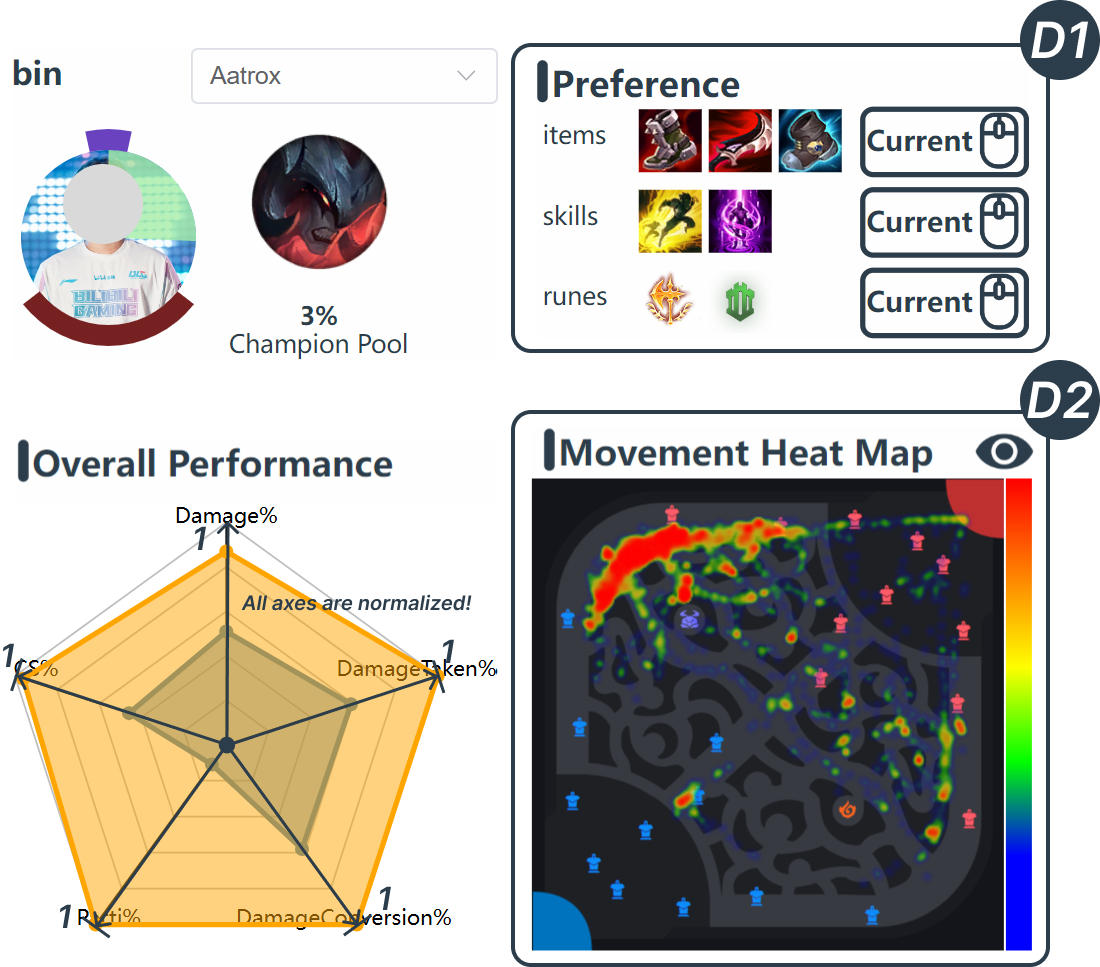}
        \caption{Overview of Player Preference View: (D1) Player Loadout Configuration Preference. (D2) Movement Heatmap. }
        \label{fig:Player_Preference_View}
    \end{minipage}
\end{figure*}

 This view, as shown in \autoref{fig:Team_Performance_View}, comprised of four sub-views \textbf{Player Glyph}, \textbf{Overall Performance}, \textbf{Carry Position}, \textbf{Lineup Combo}, provides a comprehensive presentation of a team's performance in a specific event, allowing users to compare it with the performance in the current event. Easy navigation between teams and corresponding events is facilitated through the dropdown box.

\par \textbf{Player Glyph.} The player glyph(\autoref{fig:Team_Performance_View}(C1)) enables users to swiftly view and compare the performance of each player in a team during a specific event. Various aspects are represented within the glyph. The Player Glyph's alternative design and additional details can be found in \autoref{appendix:alt_playerGlyph}.
\begin{itemize}
    \item The arc above the player's avatar reflects the damage taken ratio, with \textbf{E2} highlighting the significance of this metric in measuring a tank's performance.
    \item The arc below the player's avatar signifies the damage conversion rate. \textbf{E2} emphasizes the importance of this metric, particularly for the team's core member, who usually holds the highest economy. A higher damage conversion rate enhances their ability to maximize economic benefits.
    \item The sector covering the top of the player's avatar encodes the player's KDA in that event.
\end{itemize}



\par \textbf{Overall Performance.} In the overall performance (\autoref{fig:Team_Performance_View}(C2)), we employ a radar plot to showcase six dimensions of the team's performance in the field. The steelblue radar represents historical performance, while the orange radar represents the current match performance. The six axes (clockwise from the top) depict normalized results of key metrics: \textit{average damage per game}, \textit{average economy per game}, \textit{average damage taken per game}, \textit{first blood rate}, \textit{average tower pushes per game}, and \textit{resource control rate}. This visualization allows users to intuitively grasp the team's ranking performance among all participating teams in that event. Importantly, these metrics are normalized—for instance, a first blood rate of 1 means the highest among all teams, not securing first blood in every match. The same encoding is employed in the player radar plot of \autoref{sec:Player_Preference_View}, featuring five axes (clockwise from the top): \textit{average damage ratio per minute}, \textit{average damage taken ratio per minute}, \textit{damage conversion rate}, \textit{team fight participation rate}, and \textit{average creep score per minute}. This consistent representation ensures a seamless and unified understanding of team and player performances across different views.


\par \textbf{Carry Position.} The \textit{Carry Position} (\autoref{fig:Team_Performance_View}(C3)) serves as a valuable tool to comprehend the team's overall strategy style and the distribution of carry positions. As stated in \textbf{E3}, ``\textit{Economy reflects which position is the carry better than KDA, as KDA may have some randomness, while economy is usually deliberately allocated.}'' Therefore, the economic situation of each position is normalized to serve as the \textit{carry\_score}. Meanwhile, \textit{aggro\_label} is derived based on features such as KDA, drakes killed, turret pushed, and team economy through K-means~\cite{hartigan1979k} clustering, with the distance to the cluster center quantified as the \textit{aggro\_score}. For the visualization design, in the pentagram chart, every dot represents a victorious match, with green indicating $aggro\_label = 0$, and red indicating $aggro\_label = 1$. The dot's position is determined by the main core position of the lineup. For instance, if the primary carry position is mid, the dot falls in the mid direction. The distance of the dot from the center is dictated by the \textit{aggro\_score}, with higher scores positioned farther out. Hovering over a dot reveals its value and clicking unveils the corresponding pentagram chart, allowing users to hover over the central pentagon and examine specific data. Within the pentagram chart, each pointer signifies a champion position, and the length of the pointer denotes the \textit{carry\_score} of that position, with longer pointers indicating higher scores. This interactive visualization offers a comprehensive understanding of the team's strategy style and the different carry positions' impact on overall performance in victorious matches.

\par \textbf{Lineup Combo.} The \textbf{Combo} (\autoref{fig:Team_Performance_View}(C4)) displays the champion combinations that most frequently appeared and achieved the highest win rates for the team in a specific event. The process involves tallying all champions combinations ranging from 2 to 5 champions. Notably, combinations of 4 and 5 champions typically occurred only once, leading to the retention of combinations involving 2 and 3 champions for detailed presentation.

\subsubsection{Player Preference View}
\label{sec:Player_Preference_View}

\par This view (\autoref{fig:Player_Preference_View}) provides a comprehensive overview of a player's performance in a specific event, coupled with their historical preferences for champion selection. Users can easily switch between the player's chosen champions via a dropdown menu.



\par \textbf{Player Loadout Configuration Preference.} Experts recommended that examining a player's historical loadout preferences offers valuable insights into their playstyle, helping users better understand inconsistent behaviors during matches from an individual perspective. As depicted in \autoref{fig:Player_Preference_View}(D1), the system showcases the top three most frequently used items (equipment), skills, and runes for a specific champion in a particular event. Users can press and hold a button on the right to view the equipment, skills, and runes used by the player in the current match. This feature deepens the analysis of a player's strategic decisions and preferences, offering a more nuanced understanding of their gameplay.

\begin{figure*}[h]
  \centering
  \includegraphics[width=0.9\textwidth]{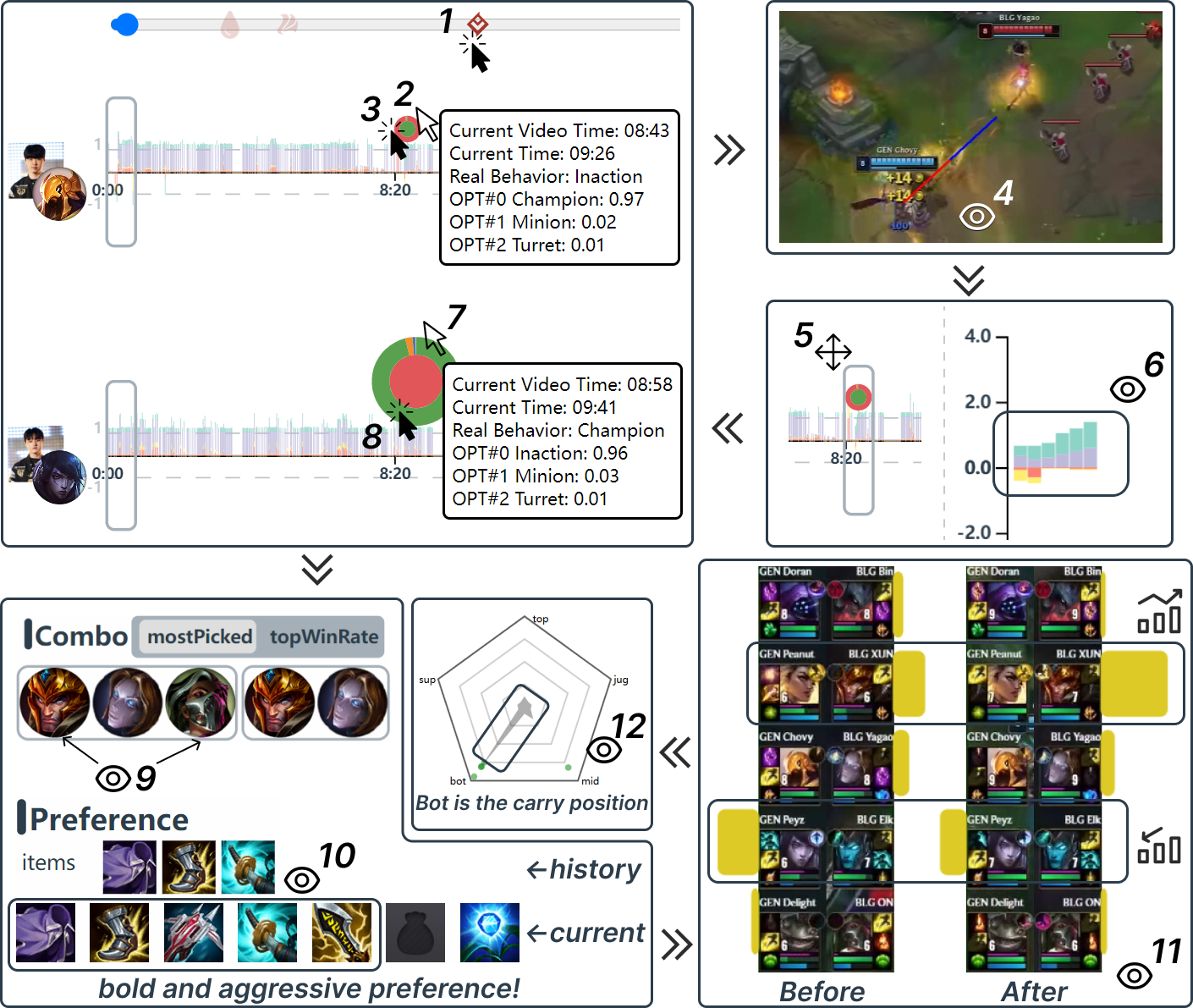}
  \caption{Event 1: (1) Identified inconsistencies in \textit{Chovy}'s performance at 09:26 and \textit{Peyz}'s at 09:41. (2) Analyzed \textit{Chovy}'s behavior, revealing a mismatch between actual ``Inaction'' and predicted ``Champion'' with 0.97 probability. (3-4) Explored the \textit{Replay View}, finding model coordinates near enemy Mid Laner \textit{Yagao}. (5) Investigated \textit{Chovy}'s actions at 9:21, observing blood/gold feedback and coordinate/behavior importance. (6) Noted negative feedback affecting behavior prediction. (7) Discovered a discrepancy in \textit{Peyz}'s actual ``Champion'' behavior versus predicted ``Inaction''. (8) Found \textit{Peyz} attacking ``Jarvan'', explaining the inconsistency. (9-10) Explored \textit{BLG}'s combo and \textit{Peyz}'s item preferences. (11) Compared real-time gold at inconsistency and post-Hextech Drake event. (12) Highlighted \textit{Peyz}'s resource advantage but aggressive actions leading to teammate \textit{Peanut}'s death and Hextech Drake loss.}
  \label{fig:case_event1}
\end{figure*}

\par \textbf{Movement Heatmap.} According to \textbf{E3}, a player's movement data reflects their playstyle and can indicate team tendencies, such as which lane the Jungler tends to gank when picking Champion A. As shown in \autoref{fig:Player_Preference_View}(D2), the system provides a coordinate heatmap that visualizes the player's movement with a specific champion in a given event. Users can press the current button to display the player's movement heatmap for the ongoing match. This feature helps users identify strategic patterns and preferences, enabling deeper analysis of gameplay dynamics.

\section{Evaluation}
\par We evaluated the effectiveness of the \textit{StratIncon Detector} in multiple ways. First, we described one case study (\textbf{RQ3, RQ5}) with our domain experts (\textbf{E2-E4}) who had participated in our user-centric design process. Second, we invited 24 participants and conducted a user study to further assess the efficacy of \textit{StratIncon Detector} (\textbf{RQ4}). Finally, we conducted expert interviews about their user experience and feedback (\textbf{RQ4, RQ5}).

\begin{figure*}[ht]
  \centering
  \includegraphics[width=0.9\textwidth]{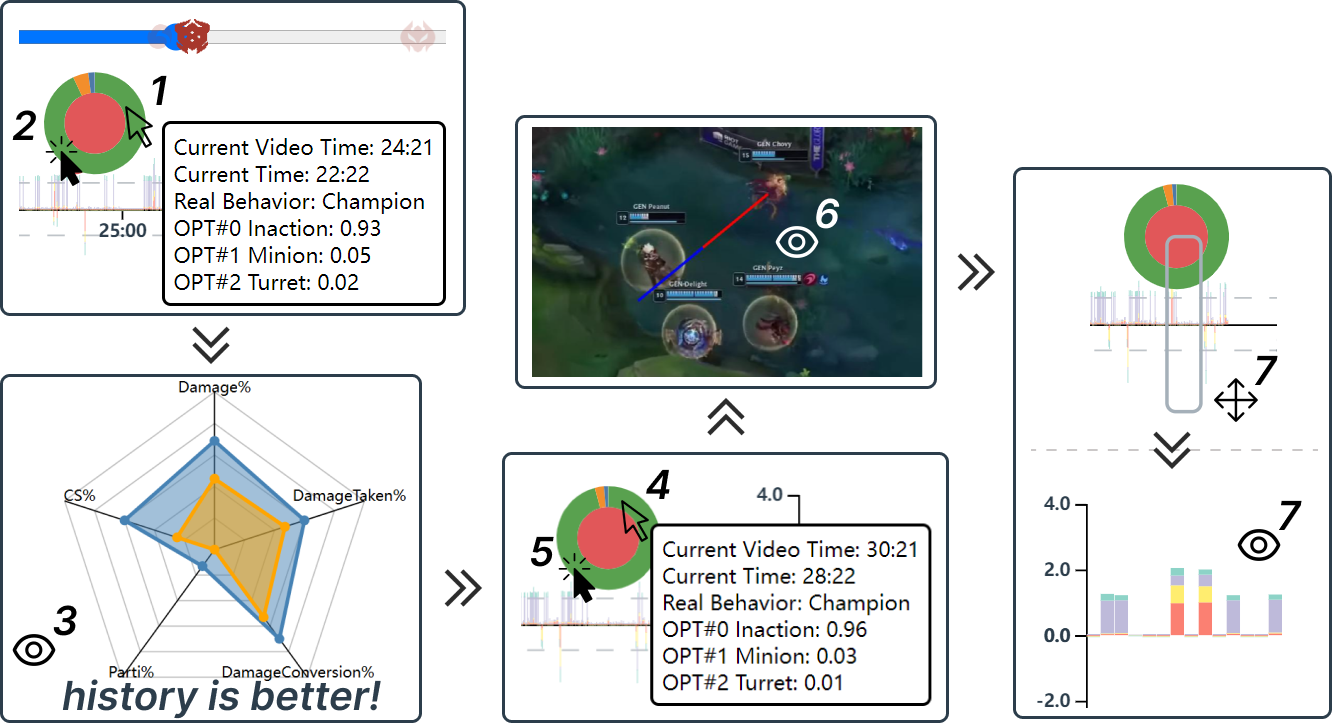}
  \caption{Event 2 \& 3: (1) Noted that \textit{Doran}'s observed behavior was ``Champion'', while the model recommended ``Inaction'' as the preferred professional choice. (2) Observed \textit{Doran} engaging the enemy's top laner despite the presence of all enemy champions. (3) Reviewed \textit{Doran}'s overall performance in the match. (4) \textit{Chovy}'s actual behavior was ``Champion'', but the model suggested ``Inaction'' as the preferred professional behavior. (5) Model predicted a position distant from enemy champions and closer to allies. (6) Significance of blood and coordinate factors notably increased. (7) Model's prediction for \textit{Chovy}'s inaction primarily relied on health and positional information.}
  \label{fig:case_event2_3}
\end{figure*}

\subsection{Case Study}
\label{sec:case study}
\par To gain a comprehensive understanding that merges theory and practice, we collaborated with experts \textbf{E2-E4} on a detailed case study of the initial match between two professional teams \textit{GEN (Gen.G)} and \textit{BLG (Bilibili Gaming)}. This study identified three pivotal turning points and analyzed key inconsistencies highlighted by the experts. We began with a 10-minute training session to familiarize them with the system. The case study focused on the first game of the best-of-five series from the \textit{S13 Worlds Championship}, which all three experts had viewed but had not thoroughly analyzed with their coaches prior.

\subsubsection{Event 1: Hextech Drake}
\label{sec:event1}
\par In the beginning, \textbf{E2} input the GameID of this match for analysis. Sequentially navigating through various events on the event timeline, \textbf{E2} focused on the Hextech Drake event. When \textbf{E2} selected the Hextech Drake event icon, they observed significant points of inconsistency for the Mid player of \textit{GEN}, \textit{Chovy}, at 09:26 and for the Bot player of \textit{GEN}, \textit{Peyz}, at 09:41 (\autoref{fig:case_event1}(1)).

\par Initially, \textbf{E3} hovered over \textit{Chovy}'s inconsistency  (\autoref{fig:case_event1}(2)) and noticed that \textit{Chovy}'s real behavior was ``Inaction'', whereas the model's predicted behavior with the highest probability was ``Champion'', reaching 0.97. This prompted \textbf{E2} to seek more context by examining the frames before and after to understand the model's prediction. Thus, \textbf{E3} clicked on the inconsistency  (\autoref{fig:case_event1}(3)), navigating to that moment in the \textit{Replay View}, and observed that the model's coordinate prediction was near the enemy Mid Laner \textit{Yagao} (\autoref{fig:case_event1}(4)). \textbf{E3} speculated, ``\textit{Chovy was playing `Azir', saw Peanut (the Jungler of GEN) nearby on the minimap and the enemy Mid Laner had Flash, so theoretically, an anticipation could be made, which helps explain the model's result.}'' However, \textbf{E2} disagreed, ``\textit{Such an anticipation is risky. The model's predicted spot is near the enemy tower's attack range, something pros usually avoid, especially with \textit{Yagao}'s healthy HP and economic lead. I believe \textit{Chovy}'s decision to remain inactive was the right call.}'' Further, \textbf{E2} dragged \textit{Chovy}'s brushbox to around 9:21 (five seconds before 9:26) (\autoref{fig:case_event1}(5)) and observed the feature importance through the magnifying glass on the right. It was noted that during these five seconds, blood and gold indeed provided negative feedback for the Champion behavior prediction, while coordinate and behavior offered high positive feedback (\autoref{fig:case_event1}(6)). \textbf{E3} further hypothesized, ``\textit{The model may have predicted these coordinates and behavior because, in the past five seconds, \textit{Chovy}'s positioning pressured \textit{Yagao}, showing constant Champion behavior, while \textit{Yagao} stayed inactive during his retreat.}''

\par Upon closer inspection, it was observed that \textit{Peyz}'s actual behavior was `champion', while the prediction model suggested `inaction' as the preferred professional behavior (\autoref{fig:case_event1}(7)). Clicking on this inconsistency point revealed that \textit{Peyz} was attacking the enemy Jungler, ``Jarvan'', at that moment (\autoref{fig:case_event1}(8)).
\par \textbf{E3}, examining \textit{BLG}'s Combo preference (\autoref{fig:case_event1}(9)), noted their most picked and top win rate champion combo. \textbf{E3} commented, ``\textit{First, in the tight space, `Jarvan' and `Renata' had plenty of control, making it tough for \textit{GEN} to dodge. Plus, this was \textit{BLG}’s go-to Combo. So, backing off was the best call, just like the model predicted.}'' \textbf{E4} added a perspective, ``\textit{It's also possible that \textit{Peyz} was taking a risk for greater gains, which aligns with his aggressive playing style.}'' This was inferred as \textbf{E4} explored \textit{Peyz}'s preference items in the player preference view on the left (\autoref{fig:case_event1}(10)), indicating his bold and aggressive style.
\par To assess the impact of the inconsistency on the game situation, \textbf{E2} compared the real-time gold bar plot (\autoref{fig:case_event1}(11)) at the moment of the inconsistency and after the Hextech Drake event in the \textit{Replay View}. The analysis revealed a widened economic gap for \textit{Peanut}, an opposing player. Exploring \textit{GEN}'s overall strategic style, \textbf{E2} found that \textit{Peyz}, as a core player of the team in this \textit{S13 World Championship}, received significant resources in this game (\autoref{fig:case_event1}(12)) but his aggressive actions resulted in his teammate \textit{Peanut}'s death while covering his retreat, and the loss of the Hextech Drake. This indirectly increased the developmental pressure, widening the economic gap against their counterparts.

\subsubsection{Event 2: Baron}
\par \textbf{E2} delved into subsequent events, focusing on the Baron event. Selecting the Baron event, \textbf{E2} identified a significant inconsistency in the Top Laner of \textit{GEN} \textit{Doran}'s behavior. Hovering over the inconsistency point at 22:22, \textbf{E2} observed that \textit{Doran}'s actual behavior was `Champion', while the model suggested `inaction' as the preferred professional behavior (\autoref{fig:case_event2_3}(1)). Clicking on this point for a deeper exploration (\autoref{fig:case_event2_3}(2)), \textbf{E2} found that \textit{Doran} was attacking the enemy's top laner while all enemy champions present. Reviewing \textit{Doran}'s overall performance in the match (\autoref{fig:case_event2_3}(3)), \textbf{E2} commented on the suboptimal conditions reflected in the radar chart, with damage, CS, participant rate, and damage conversion below historical data. \textbf{E2} remarked, ``\textit{Doran's aggressive actions were unwarranted, contributing to \textit{GEN}'s significant defeat in the team fight}''. \textbf{E2} added that, ``\textit{Indeed, \textit{Doran}'s incorrect judgment led to the team fight's collapse. Had he followed the model's prediction, the outcome of the team fight would not have been so dire.}'' In conclusion, \textit{Doran}'s mistake resulted in the loss of a team member for \textit{GEN}, ultimately leading to the forfeiture of their claim to the Baron.


\subsubsection{Event 3: Elder Dragon}
\par Upon selecting the Elder Dragon event, \textbf{E2} identified a notable inconsistency in \textit{Chovy}'s behavior. Hovering over the inconsistency point at 28:22 (\autoref{fig:case_event2_3}(4)), \textbf{E2} observed that \textit{Chovy}'s actual behavior was `Champion', but the model suggested `inaction' as the preferred professional behavior.
\par To explore the specific context of this inconsistent action, \textbf{E2} clicked on the point (\autoref{fig:case_event2_3}(5)) and noted that \textit{Chovy}'s HP was low, and his position was dangerously close to multiple enemy champions, posing a significant risk. The model predicted a position far from enemy champions and closer to allies (\autoref{fig:case_event2_3}(6)).
\par \textbf{E2} commented, ``\textit{This mistake led to \textit{Chovy} being instantly killed, disrupting the team's formation and losing the output environment, resulting in the collapse of the team fight.}'' \textbf{E3} then dragged \textit{Chovy}'s brushbox to around 28:17 (five seconds before 28:22) and observed the feature importance through the magnifying glass on the right (\autoref{fig:case_event2_3}(7)). It was found that in these five seconds, the significance of blood and coordinate increased significantly (\autoref{fig:case_event2_3}(7)). This indicates that the model's prediction for \textit{Chovy} to take inaction was primarily based on health and positional information, aligning closely with human judgment. \textbf{E3} stated that \textit{Chovy}'s misjudgment of the safe range of ``Azir'' in this team fight was the catalyst for the defeat.

\subsection{User Study}
\par To comprehensively assess the usability and effectiveness of \textit{StratIncon Detector} in identifying strategy inconsistencies (\textbf{RQ3}), we drew upon several comparative studies in CSCW, CHI and IUI research~\cite{xia2019peerlens,chen2023meetscript,liu2023bpcoach,liu2023coargue,liu2024biaseye}. Building on these references, we designed an IRB-approved between-subjects user study to explore the inconsistent behaviors of players in professional matches.

\subsubsection{Participants}
\par We recruited $24$ participants ($22$ males, $2$ females), aged $18$ to $28$ years (mean = $21.46$, SD = $2.43$). Given the system's complexity, participants were selected for their deep understanding of relevant games and competitive events, including players, coaches, and analysts from local university teams. Despite their youth, $58.33\%$ had over $3$ years of esports experience, and $12.5\%$ had more than $5$ years. Their professional expertise was crucial for validating the system's accuracy and performance, ensuring our findings are effective and practically relevant for future optimizations.

\subsubsection{Procedure}
\par Before the experiment, participants watched a 30-minute match video that they would analyze, ensuring none had prior experience with it. To avoid order effects, we implemented a between-subjects design, dividing the $24$ participants—$22$ males and $2$ females—into two groups of $12$. Group A used our system, \textit{StratIncon Detector}, while Group B analyzed the match using \textit{scoregg} and \textit{LOL Esports}, common tools for professional post-match analysis. All participants, including coaches and analysts, focused on identifying inconsistencies in player behavior and completing designated tasks. Prior to the analysis, participants attended a 10-minute orientation session to familiarize themselves with the system's content, data distribution, and operations, allowing for preliminary trials. They then had 30 minutes to complete the tasks (\autoref{tab:user_task}), which were validated in a pilot experiment. Upon finishing, participants filled out a survey using a 7-point Likert scale to assess their experiences and perceptions of the two analysis tools. They were also encouraged to share their feedback and insights.

\begin{table}[h]
    \caption{The types of tasks in the user experiment.}
    \label{tab:user_task}
    \centering
    \scalebox{0.8}{
    \begin{tabular}{llcc}
    \toprule
    ID & ~ & Task \\
    \midrule
    T1& ~ & Explore a player's performance in this match\\
    T2& ~ & Investigate the positioning of a player during this match \\
    T3& ~ & Identify whether a team is aggressive and determine the carry position \\
    T4& ~ & Identify any inconsistent behavior in this match and analyze its reason\\
    \bottomrule
    \end{tabular}}
\end{table}

\begin{table*}[h]
    \caption{The user study questionnaire structure and comparative nonparametric-test results between the \textit{StratIncon Detector} and the baseline system across \textit{Insightfulness}, \textit{System Performance}, \textit{User Experience \& Usability}, and \textit{Team Collaboration} (*: $p<0.05$, **: $p<0.01$).}
    \label{tab:userStudy}
    \centering
    \resizebox{\textwidth}{!}{
    \begin{tabular}{llllll}
    \toprule
    Problem & Theme & Criteria & Content & $u$-value & $p$-value\\
    \midrule
    P1 & \multirow{4}{*}{Insightfulness} & Interprebility & I can explain inconsistent behavior. & 36 & 0.028*\\
    P2 & ~ & Effectiveness & I can analyze how inconsistent behaviors affect the game. & 20.5 & 0.002**\\
    P3 & ~ & Comparability & I can compare a player's current and past performances. & 24 & 0.004**\\
    P4 & ~ & Comprehensiveness & The system comprehensively tracks players' inconsistencies. & 37.5 & 0.034*\\
    
    \midrule
    P5 & \multirow{3}{*}{System Performance} & Accuracy & I can accurately find inconsistencies in player behavior. & 17.5 & 0.001**\\
    P6 & ~ & Efficiency & The system quickly identifies players' inconsistencies. & 27.5 & 0.007**\\
    P7 & ~ & Clarity & The system's data display is clear enough. & 47.5 & 0.128\\
    
    \midrule
    P8 & \multirow{2}{*}{UX \& Usability} & Simplicity & The system is easy to learn and use. & 46.5 & 0.129\\
    P9 & ~ & Intelligibility & The information provided is easy to understand. & 51 & 0.173\\

    \midrule
    P10 & \multirow{2}{*}{Team Collaboration} & Supportiveness & I can analyze key support actions between players. & 24.5 & 0.004**\\
    P11 & ~ & Strategic Execution & I can analyze the team's strategic execution. & 26 & 0.006**\\
    
    \bottomrule
    \end{tabular}}
\end{table*}

\begin{figure*}[h]
  \centering
  \includegraphics[width=\textwidth]{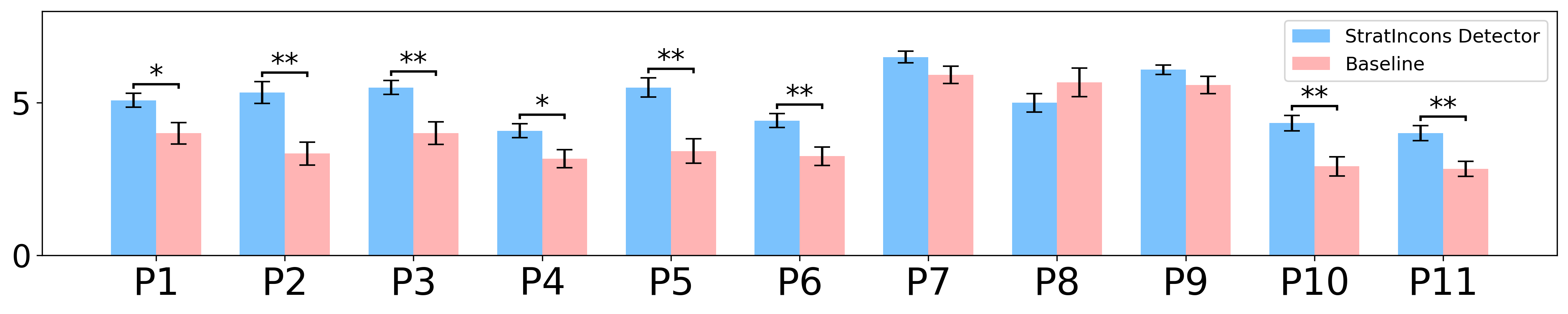}
  \caption{The ratings between \textit{StratIncon Detector} and the baseline system for P1-P10, corresponding to \textit{Interprebility}, \textit{Effectiveness}, \textit{Comparability}, \textit{Comprehensiveness}, \textit{Accuracy}, \textit{Efficiency}, \textit{Clarity}, \textit{Simplicity}, \textit{Intelligibility}, \textit{Supportiveness}, and \textit{Strategic Execution}, respectively.}
  \label{fig:user_study_evaluation}
\end{figure*}
\subsubsection{Hypothesis}
\par Since the samples from all participants did not adhere to a normal distribution, and considering the relatively small sample size in our experiment, we utilized the \textit{Mann-Whitney U Test}~\cite{mcknight2010mann}. This non-parametric test is suitable for comparing median differences between two independent samples. We formulated the following hypotheses:
\begin{itemize}
    \item \textbf{H1}: \textit{StratIncon Detector} surpasses the baseline system in terms of \textit{Interpretability} \textbf{(H1a)}, \textit{Effectiveness} \textbf{(H1b)}, \textit{Comparability} \textbf{(H1c)}, and \textit{Comprehensiveness} \textbf{(H1d)}.
    \item \textbf{H2}: The \textit{StratIncon Detector} outperforms the baseline system in \textit{Accuracy} \textbf{(H2a)}, \textit{Efficiency} \textbf{(H2b)}, and \textit{Clarity} \textbf{(H2c)}.
    \item \textbf{H3}: The \textit{StratIncon Detector} excels over the baseline system in terms of \textit{Simplicity} \textbf{(H3a)} and \textit{Intelligibility} \textbf{(H3b)}.
    \item \textbf{H4}: The \textit{StratIncon Detector} shows better performance than the baseline system in terms of \textit{Supportiveness} \textbf{(H4a)} and \textit{Strategic Execution} \textbf{(H4b)}.
\end{itemize}

\subsubsection{Results}
\par We performed \textit{Mann-Whitney U Test} to assess the difference in the participants' ratings and the results are shown in \autoref{tab:userStudy} and \autoref{fig:user_study_evaluation}.

\par \textbf{Insightfulness.} In comparison to the baseline system, the \textit{StratIncon Detector} demonstrated higher scores in \textit{Interpretability}, \textit{Effectiveness}, \textit{Comparability}, and \textit{Comprehensiveness}, suggesting that it provides more satisfactory information regarding players' inconsistent behaviors. When evaluating \textit{Interpretability}, participants noted that using the \textit{StratIncon Detector} (Mean = 5.083, SD = 0.793, Median = 5) facilitated a better explanation of the reasons behind players' inconsistent behaviors compared to the baseline system (Mean = 4, SD = 1.206, Median = 4.5) (\textbf{H1a} supported, $p=0.028$*). Additionally, participants believed that the \textit{StratIncon Detector} (Mean = 5.333, SD = 1.231, Median = 5) provided a better understanding of the impact of players' inconsistent behaviors on the subsequent course of the game than the baseline system (Mean = 3.333, SD = 1.303, Median = 3) (\textbf{H1b} supported, $p=0.002$**). Regarding \textit{Comparability}, participants generally felt that the \textit{StratIncon Detector} (Mean = 5.5, SD = 0.798, Median = 5.5) provided a more comprehensive comparison of historical and current performances than the baseline system (Mean = 4, SD = 1.279, Median = 4) (\textbf{H1c} supported, $p=0.004$**). P4 (ID:4, male, age: 27, role: coach) highlighted, ``\textit{The \textit{StratIncon Detector} stands out in comparing a player's or team's historical performances with their current performance, offering a wide coverage of information.}'' In terms of the \textit{Comprehensiveness} of inconsistencies, participants generally viewed the \textit{StratIncon Detector} (Mean = 4.083, SD = 0.793, Median = 4) as superior to the baseline system (Mean = 3.167, SD = 1.03, Median = 3.5) (\textbf{H1d} supported, $p=0.034$*).

\par \textbf{System Performance.} The \textit{StratIncon Detector} (Mean = 5.5, SD = 1.087, Median = 5) received more favorable reviews than the baseline system (Mean = 3.417, SD = 1.379, Median = 4) in terms of the \textit{accuracy} in displaying players' inconsistent behaviors (\textbf{H2a} supported, $p=0.001$**). Additionally, the \textit{StratIncon Detector} (Mean = 4.417, SD = 0.793, Median = 4) received better feedback for the efficiency of identifying inconsistencies compared to the baseline system (Mean = 3.25, SD = 1.055, Median = 3) (\textbf{H2b} supported, $p=0.007$**). P1 (ID: 1, male, age: 18, role: player) stated, ``\textit{When I review matches I've played in, analyzing inconsistent behaviors typically involves going through the game replay. By the time I find the key event, rewatch actions from different perspectives, and think about possible mistakes, a lot of time has already passed. But with the \textit{StratIncon Detector}, I can quickly identify each player's inconsistent behaviors by clicking on the points of inconsistency, which is really helpful.}'' Additionally, both the \textit{StratIncon Detector} (Mean = 6.5, SD = 0.674, Median = 7) and the baseline system (Mean = 5.917, SD = 0.996, Median = 6) provided very clear data. Despite the \textit{StratIncon Detector}  performing better, the difference was not significant (\textbf{H2c} rejected, $p=0.128$).

\par \textbf{User Experience \& Usability.} Both systems performed well in terms of \textit{Simplicity} and \textit{Intelligibility}. The simplicity score of the baseline system (Mean = 5.667, SD = 1.614, Median = 6) was slightly higher than that of the \textit{StratIncon Detector} (Mean = 5, SD = 1.044, Median = 5), but the difference was not significant (\textbf{H3a} rejected, $p=0.129$). Additionally, participants found both systems easy to understand, resulting in high \textit{intelligibility} scores for the proposed system (Mean = 6.083, SD = 0.515, Median = 6) and the baseline system (Mean = 5.583, SD = 0.996, Median = 6), with no significant difference between them (\textbf{H3b} rejected, $p=0.173$). P6 (ID: 6, male, age: 22, role: player) commented, ``\textit{Using \textit{StratIncon Detector} was a great overall experience, especially because it was so easy to get started with.}''

\par \textbf{Team Collaboration.} The \textit{StratIncon Detector} (Mean = 4.333, SD = 0.888, Median = 4) significantly outperformed the baseline system (Mean = 2.917, SD = 1.084, Median = 3) in terms of \textit{Supportiveness}, indicating its greater efficacy in assisting users in accessing and analyzing players' support actions during critical moments (\textbf{H4a} supported, $p=0.004$**). P22 (ID: 22, age: 20, role: player) remarked, ``\textit{With the \textit{StratIncon Detector}, I can simultaneously observe if all players are acting inconsistently at any given moment and quickly switch to that moment to evaluate each player's supportive behavior. The teamwork issues revealed can, in turn, foster improved team collaboration in our next engagement.}'' Regarding \textit{Strategic Execution}, the \textit{StratIncon Detector} (Mean = 4, SD = 0.853, Median = 4) also demonstrated superior performance compared to the baseline system (Mean = 2.833, SD = 0.835, Median = 3), showing its ability to comprehensively analyze team strategic execution (\textbf{H4b} supported, $p=0.006$**). P24 (ID: 24, age: 22, role: player) noted, ``\textit{When using the \textit{StratIncon Detector}, I can discern historical preferences for aggressive playstyles and resource allocation biases through the left view. Combining this with the inconsistencies displayed in the right view helps me analyze whether the team strategy is being executed correctly and identify the crux of team collaboration issues.}''

\subsection{Expert Interview}
\par We conducted a semi-structured interview with experts \textbf{E1-E4} to gather their feedback. The experts noted that the system's accuracy in detecting inconsistencies is impressively high, with many predictions aligning closely with their expectations. It not only identifies surface-level inconsistencies but also reveals additional details, such as illogical resource deployment in the \textit{GEN} team, including impractical core position distribution. \textbf{E2} praised the system's interpretability, highlighting how the historical performance view and real-time gold bar plot aid in understanding the origins and implications of inconsistencies. All four experts agreed that the \textit{StratIncon Detector} offers valuable insights. \textbf{E3} remarked on the intuitive design of the \textit{team performance view} and \textit{player preference view}, noting their user-friendly visualizations. \textbf{E4} emphasized the \textit{player inconsistency view}, stating, ``\textit{It is immensely beneficial, offering predicted strategies while quickly identifying all points of inconsistency, significantly streamlining our review process.}'' \textbf{E2} further elaborated on the system's ability to detect subtle inconsistencies, asserting it uncovers a broader range than manual analysis typically allows, focusing on often-overlooked details. He also emphasized the contributions of the \textit{team performance view} and \textit{player preference view} in providing insights into a player's style and the roots of certain inconsistencies.

\section{Discussion and Limitation}

\subsection{Contributions to Strategy Analysis}
\par The \textit{StratIncon Detector} highlights behavioral differences among teams and players, enabling professional teams to better understand their opponents' tactics and strategies. This insight aids in correcting strategic deficiencies and prepares teams for diverse opponents and scenarios. Moreover, \textbf{E3} and \textbf{E4} emphasized that the system enhances players' self-awareness of their strengths and weaknesses during matches, leading to more informed and strategic decisions in future games.
\par In the case study interview, \textbf{E2} remarked, ``\textit{Using this system, we quickly identified all points of inconsistency, facilitating multiple rounds of explanation and deduction.}'' The case study, detailed in \autoref{sec:case study}, illustrated how experts employed the \textit{StratIncon Detector} to analyze inconsistent behaviors and articulate their thought processes iteratively. The system first presented all inconsistencies alongside predicted actions and their impacts. Experts then integrated feature contributions, original video data, and historical performance to refine or reject these predictions. Overall, the system streamlined the analysis process, providing an effective entry point for reviews and improving efficiency. Its ``human-in-the-loop'' design allows experts to focus on the review itself—explaining predicted actions and analyzing the effects of current behaviors—rather than merely processing black-box outputs, thus enhancing review quality.

\subsection{Contributions to Team Construction}
\subsubsection{Shifting the Focus of Communication to Reduce Conflicts.} As shown in \prefix{\texttt{\textbf{[F4]}}}, players have inherent subjective biases and assumptions that lead to different interpretations of strategies, resulting in conflicts. \textbf{E3} also expressed in the interview during the Observational Study (\autoref{sec:observational study}), ``\textit{During post-game reviews, there are often disputes and arguments among team members regarding what the proper action should have been. As the shotcaller, when these disagreements occur frequently, I start to wonder if someone is deliberately not cooperating or opposing me.}'' Therefore, frequent cognitive discrepancies among teammates can lead to a decrease in trust and disrupt team harmony. In the case study of Event1 (\autoref{sec:event1}), \textbf{E2} and \textbf{E3} had disagreements, but ultimately, \textbf{E3} was persuaded by the system's feature importance and \textbf{E2}'s interpretation of the model's prediction coordinates. This process is in stark contrast to the heated arguments observed in the Contextual Inquiry (\autoref{sec:Contextual Inquiry}). The \textit{StratIncon Detector} displays inconsistent behaviors and their impacts, providing additional information to help experts understand or correct this predicted behavior, thus facilitating communication among teammates about their abstract ``understanding of the game''. Additionally, when the preferred professional behavior can be roughly and objectively determined, it allows team members to focus more on explaining the actions and analyzing the impact of current behaviors, rather than simply blaming each other and questioning ``\textit{I asked you to clear the minions, but why did you go for the jungle instead?}''


\subsubsection{Engaging Everyone in the Analysis to Enhance Team Performance}
\par Traditional review tools often limit discussions to just one or two critical team fights due to efficiency constraints, as noted in \prefix{\texttt{\textbf{[F5]}}}. In contrast, the \textit{StratIncon Detector} offers a comprehensive view of each player's inconsistent behaviors. \textbf{E2} highlighted this advantage after the case study, ``\textit{In past reviews, we could pinpoint the `culprit' behind our loss, but usually only that player learned from it. With the \textit{StratIncon Detector}, everyone can see their own mistakes and has room for improvement.}'' This perspective was also supported by experts' analyses of multiple players in the case study. The data-driven approach of the \textit{StratIncon Detector} offers a level of comprehensiveness that traditional tools lack, encouraging all team members to reflect on their inconsistencies. Consequently, each review fosters improvements in overall team capabilities and cohesion.

\subsection{Generalizability and Scalability}
\label{sec:6.2}
\par In terms of generalizability, experts suggest that our system has strong potential for adaptation across different MOBA games, provided that data preprocessing is handled carefully. This adaptability enables the system to capture a wider range of behavioral dynamics with appropriate data preparation. When log data is unavailable, involving domain experts in defining behaviors could be beneficial. Regarding scalability, the current implementation of the \textit{StratIncon Detector} utilizes a dataset of approximately 100,000 frame images, with each match generating 40-50 inconsistency points—around 10 per player. These inconsistencies tend to cluster during critical gameplay phases like team battles. To improve this, future updates could integrate tree-based visualizations~\cite{sun2016videoforest} to reduce visual clutter in high-level summaries, or implement filtering mechanisms to focus on specific levels of inconsistency for more targeted analysis.

\subsection{Limitations and Future Work}
\par Our work has several limitations. \textbf{First}, the training data comes from only 50 matches across six events, which does not adequately represent the diversity of teams and player styles. While these matches yield over 100,000 frame images, simply increasing the match count would significantly heighten the training burden. Moreover, match videos from the \textit{LOL Esports} website exhibit frame-skipping phenomena, characterized by abrupt transitions or sparse frames. While this may occasionally cause alignment issues, the overall impact on our system's performance remains acceptable, as a behavior typically spans multiple frames. Future research should explore balancing matches quantity with frame selection. \textbf{Second}, while ideal system evaluation would involve experts and participants replaying their matches for comparative analysis, our user study participants are drawn from various teams and roles to enhance statistical reliability. Additionally, the collaborating experts lack prominence, and their match records are poorly documented, limiting the data available for robust model training and affecting statistical insights. Nevertheless, sufficient match data could greatly benefit players, as outlined in \autoref{sec:6.2}. Future efforts should focus on deeper collaboration with more esports teams and developing a systematic feedback mechanism for continuous data flow. \textbf{Third}, the experts' predefined behavior still lacks precision. However, in the absence of detailed log data, like skill release direction and skill target, coarse-grained expert definitions remain currently the most reliable approach. \textbf{Lastly}, experts also noted some additional potentially lower-ranked factors beyond the primary concerns in \prefix{\texttt{\textbf{[KC1-KC3]}}}, such as skill cooldown durations, minion waves and hero positioning. Although these factors may seem minor, \textbf{E4} emphasized their strategic importance: ``\textit{Cooldowns might seem less crucial late in the game, but we often emphasize `wait for cd!' to ensure engagements are optimally timed.}'' This highlights the critical role of timing and ability management in gameplay. Furthermore, battlefield environmental factors such as bushes and vision mechanics should also be considered. Expanding these considerations will require more comprehensive log data and map layout data, posing a challenge for future research.

\section{Conclusion}
\par This study examines the dynamics of MOBA games, emphasizing the balance between strategic planning and real-time decision-making in professional esports. We introduce the \textit{StratIncon Detector}, a visual analytics system that helps players and coaches analyze strategy inconsistencies in matches. Through a case study, a user study with 24 participants, and expert interviews, we showcase the system's effectiveness in areas such as \textit{Insightfulness}, \textit{System Performance}, and \textit{Team Collaboration}. The findings reveal that the \textit{StratIncon Detector} enables users to identify inconsistencies, understand their causes, and assess their impacts on game outcomes, ultimately enhancing team collaboration in esports.


\begin{acks}
\par We thank anonymous reviewers for their valuable feedback. This work is supported by grants from the National Natural Science Foundation of China (No. 62372298), Shanghai Engineering Research Center of Intelligent Vision and Imaging, Shanghai Frontiers Science Center of Human-centered Artificial Intelligence (ShangHAI), and MoE Key Laboratory of Intelligent Perception and Human-Machine Collaboration (KLIP-HuMaCo).
\end{acks}

\bibliographystyle{ACM-Reference-Format}
\bibliography{sample-base}

\appendix

\section{Details of All Five Behaviors}
\label{appendix:behaviors}
\par The following paragraphs introduce each type of events:
\begin{itemize} 
    \item \textbf{Minion:} Players clear minions, which spawn from both teams' bases, to gain resources and push waves toward enemy turrets. A player is classified as engaging in ``Minion'' behavior when attacking minions in the current frame.
    \item \textbf{Champion:} Players attack enemy champions to disrupt their control, using tactics like direct combat or ambushes. A player is considered engaging in ``Champion'' behavior when using skills or basic attacks against enemy champions.
    \item \textbf{Resource:} Players secure large neutral objectives (e.g., Epic Monsters) for team-wide benefits like damage boosts or healing. A player is engaging in ``Resource'' behavior when targeting these neutral objectives.
    \item \textbf{Turret:} Destroying enemy turrets expands map control and opens access to the enemy's nexus. A player is classified as engaging in ``Turret'' behavior when attacking enemy turrets.
    \item \textbf{Inaction:} Players may remain inactive to avoid risks or prepare for battles. Inaction is recorded when a player is not engaged in any of the above behaviors in the current frame.
\end{itemize}


\section{Design Alternative of the Inconsistency Detector}
\label{appendix:alt_inconsDetector}
\par
\begin{figure}[h]
  \centering
    \includegraphics[width=0.9\linewidth]{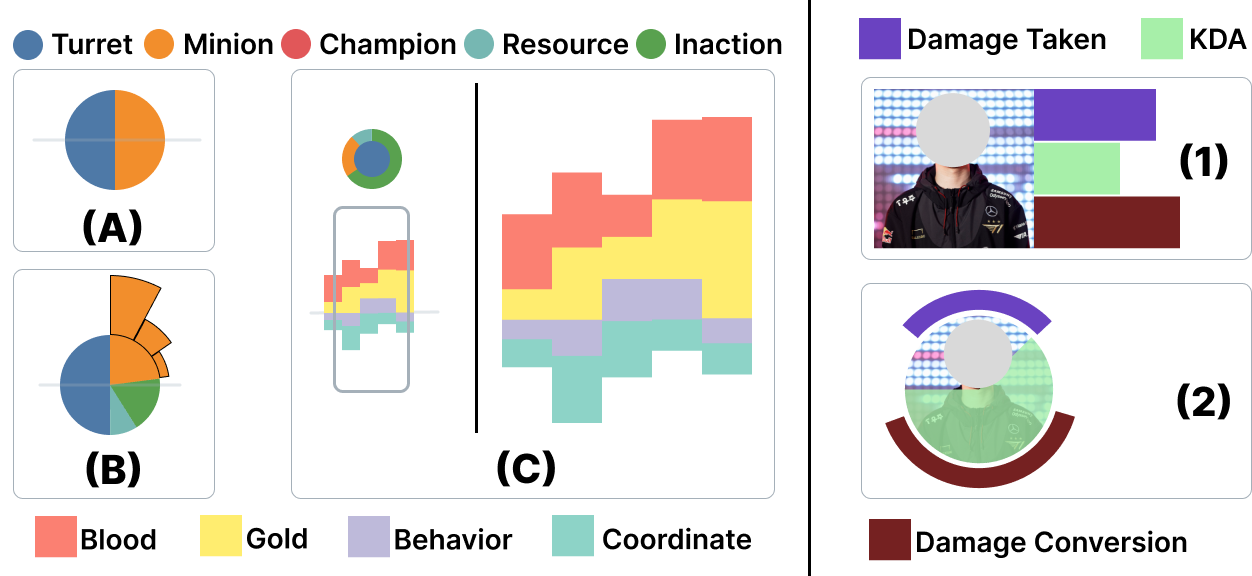}
    \caption{Design Alternative for Inconsistency Detector(A-C) and Player glyph(1-2).}
    \label{fig:design_alternative}
\end{figure}

 Through expert consultations, he \textbf{Inconsistency Detector} underwent three design iterations. Initially, depicted in \autoref{fig:design_alternative}(A), it featured a simple pie chart with two semi-circles: the left half representing red behavior and the right half showing the most probable preferred professional behavior. Feedback from \textbf{E1} highlighted a significant shortcoming: the absence of essential feature importance data, which is crucial for enhancing the interpretability of the preferred professional strategy. This feedback led to the development of \autoref{fig:design_alternative}(B). In this iteration, we displayed the top three preferred professional behaviors using a sunburst chart to clarify the feature importance corresponding to the predicted preferred professional behavior. Another crucial aspect was raised by \textbf{E2}: the need to track not only the feature importance at specific inconsistency points but also the overall trend of all significant features. To address this, we refined the design to its final form, as illustrated in \autoref{fig:design_alternative}(C). This version introduces a stacked bar plot to visualize dynamic feature importance trends. Additionally, we refined the glyph design, transitioning from two semi-circles to an inner circle and outer ring arrangement. This modification preserves the original information while avoiding confusion with pie charts. Lastly, to manage potential clutter from thousands of frame data, a brushbox tool was introduced. This feature enables users to select and zoom into specific data segments, projecting them onto a dedicated magnifying area for detailed analysis.

\section{Design Alternative of the Player Glyph}
\label{appendix:alt_playerGlyph}
\par The initial design of the player glyph is presented in \autoref{fig:design_alternative}(1), featuring a bar plot adjacent to the avatar, representing three key metrics: \textit{Damage Taken Ratio}, \textit{Damage Conversion Rate}, and \textit{KDA}. \textbf{E1} highlighted potential visual ambiguity in the current bar plot layout. While it facilitates quick metric comparisons for an individual player, it proves cumbersome for comparing different players. To address this, we iteratively refined the design, resulting in the format shown in \autoref{fig:design_alternative}(2). In this revised version, all three metrics have been reconceptualized to allow for more straightforward and intuitive horizontal comparisons.

\section{Fundamentals of MOBA Games}
\label{appendix:moba_fundamental}

\subsection{Terminology}
\par This subsection only lists the terminology\footnote{Note that some of the terminology is not officially defined by the game developers but is rather a convention commonly used by players.} relevant to this study\footnote{More terminology and jargon of LOL can be found on:\url{https://leagueoflegends.fandom.com/wiki/Terminology_(League_of_Legends)} and \url{https://www.lolguides.com/glossary}}.

\begin{itemize} 
    \item \textbf{Buff:} Temporary or permanent enhancements granted by killing neutral monsters, boosting champion stats or abilities.
    \item \textbf{Carry:} A champion type whose primary goal is to deal loads of damage. There are AD and AP carries. 
    \item \textbf{Cool Down (CD):} Cool Down (CD) means for an ability to be unavailable for a certain duration.
    
    \item \textbf{Epic Monsters:} Powerful neutral monsters that spawn at various points on the map, each providing significant advantages when defeated.
        \begin{itemize}
            \item \textbf{Baron Nashor / Baron:} Spawns in the later stages of the game. Defeating it grants the Baron Buff, which boosts minions and champions, aiding in pushing lanes and winning team fights.
            \item \textbf{Drake:} Spawns at different points in the game. Each type provides unique buffs. For example, Ocean Drake grants health regeneration, and its soul provides sustained healing for champions in combat.
            \item \textbf{Elder Dragon:} A rare monster in the later stages. Killing it grants the Elder Dragon Buff, enhancing basic attacks and abilities with percentage health damage and providing a team-wide power boost.
            \item \textbf{Rift Herald:} Found in the early-to-mid game. Defeating it drops an item that summons the Herald to lanes, helping to push waves and destroy turrets, providing a strategic advantage.
        \end{itemize}
    Baron and Rift Herald spawn in the \textbf{Baron Pit}, while Drake and Elder Dragon spawn in the \textbf{Dragon Pit}.
    \item \textbf{Gank:} To ambush an enemy with the intent of killing them. It is often the Jungler's job to gank during the laning phase to kill enemies.
    \item \textbf{Gold:} The in-game currency used to buy items.
    \item \textbf{Health Points (HP):} Health Points (or Hit Points) is the life points on a champion, minion, or turret.
    \item \textbf{The E+Q Combo of ``Jarvan IV'':} A combination of E ability (Demacian Standard) and Q ability (Dragon Strike). In this combo, ``Jarvan'' places his E at a target location and follows with Q, dashing to the standard, dealing damage, and knocking up enemies. This combo also lets him pass through impassable terrain, making it effective for engaging or escaping, with strong initiation and crowd control.
    \item \textbf{Lane:} The three paths minions follow to reach the enemy's nexus: Top Lane, Mid Lane, and Bot Lane, each assigned to a specific player: Top Laner, Mid Laner, and Bot Laner.
    \item \textbf{Jungle: } The area between lanes and the river, home to mobs that grant gold, experience, and blue/red buffs. Mobs respawn after being killed and scale with game progression.
    \item \textbf{Mana:} A resource for casting spells and abilities.
    \item \textbf{Nexus:} The core of each team’s base; destroying the enemy Nexus wins the game. It also spawns minions.
    \item \textbf{Pick:} Isolating and killing a single champion in-game to create a numbers advantage or remove a priority target.
    \item \textbf{Poke:} To damage enemies from afar with minimal risk. A tactic used by champions with long-range abilities to weaken the enemy before a fight.
    \item \textbf{Skill Combo:} Using two or more abilities in quick succession to maximize damage, crowd control, or effects.
\end{itemize}

\subsection{Evaluation Metrics}
\par MOBA games typically use a variety of statistical metrics to evaluate the performance of teams, players, and champions during different stages of a tournament. The official game platform\footnote{\url{https://lpl.qq.com/web202301/data-index.shtml}} and third-party esports data platforms\footnote{\url{https://www.scoregg.com/}} provide a wealth of statistical metrics. Based on the data from these platforms, the study will focus on the following metrics:
\par \textbf{For teams in a specific tournament}, the following six metrics will be analyzed:
\begin{itemize} 
    \item \textbf{average damage per game:} The team's total damage per game, averaged across all tournament games, reflecting overall offensive contribution.
    \item \textbf{average economy per game:} The average gold earned per game by the team, reflecting economic management.
    \item \textbf{average damage taken per game:} The team's total damage taken per game, averaged across all tournament games, reflecting defensive performance.
    \item \textbf{first blood rate:} Percentage of games where the team secured the first kill, reflecting their ability to start strong.
    \item \textbf{average tower pushes per game:} Average number of towers destroyed per game, showing the team's map control and pressure.
    \item \textbf{resource control rate:} The percentage of key objectives (such as drakes, Baron) successfully secured by the team in relation to the total number of such objectives contested.
\end{itemize}

\par \textbf{For players using a specific champion}, the following five metrics will be examined:
\begin{itemize}
    \item \textbf{average damage ratio per minute:} Similar to "average damage per game," but measured per minute.
    \item \textbf{average damage taken ratio per minute:} Similar to "average damage taken per game," but measured per minute.
    \item \textbf{damage conversion rate:} The percentage of total damage dealt to enemy champions out of the player’s overall damage, indicating focus on impactful damage.
    \item \textbf{team fight participation rate:} The percentage of team fights the player participated in, showing their involvement in key engagements.
    \item \textbf{average creep score per minute:} The average number of minions killed per minute, showing the developing efficiency.
\end{itemize}

\end{document}